\documentclass[aip,reprint,nofootinbib]{revtex4-1}
\usepackage{graphicx}
\usepackage{bm}

\begin{document}
\title{About the Use of Real Dirac Matrices in 2-dimensional Coupled Linear Optics}
\date{\today}
\author{C. Baumgarten}
\affiliation{Paul Scherrer Institute, Switzerland}
\email{christian.baumgarten@psi.ch}

\def\begeq{\begin{equation}}
\def\endeq{\end{equation}}
\def\begary{\begeq\begin{array}}
\def\endary{\end{array}\endeq}
\def\bmtx{\left(\begin{array}}
\def\emtx{\end{array}\right)}
\def\eps{\varepsilon}
\def\d{\partial}
\def\y{\gamma}
\def\w{\omega}
\def\W{\Omega}
\def\s{\sigma}
\def\ket#1{\left|\,#1\,\right>}
\def\bra#1{\left<\,#1\,\right|}
\def\bracket#1#2{\left<\,#1\,\vert\,#2\,\right>}
\def\erw#1{\left<\,#1\,\right>}

\def\Exp#1{\exp\left(#1\right)}
\def\Log#1{\ln\left(#1\right)}
\def\Sinh#1{\sinh\left(#1\right)}
\def\Sin#1{\sin\left(#1\right)}
\def\Tanh#1{\tanh\left(#1\right)}
\def\Tan#1{\tan\left(#1\right)}
\def\Cos#1{\cos\left(#1\right)}
\def\Cosh#1{\cosh\left(#1\right)}

\begin{abstract}
The Courant-Snyder theory for two-dimensional coupled linear optics is presented, based on
the systematic use of the real representation of the Dirac matrices. 
Since any real $4\times 4$-matrix can be expressed as a linear combination of these matrices,
the presented Ansatz allows for a comprehensive and complete treatment of two-dim. linear coupling.
A survey of symplectic transformations in two dimensions is presented. A subset of these 
transformations is shown to be identical to rotations and Lorentz boosts in Minkowski space-time.
The transformation properties of the classical state vector are formulated and found to be
analog to those of a Dirac spinor. The equations of motion for a relativistic 
charged particle - the Lorentz force equations - are shown to be isomorph to envelope 
equations of two-dimensional linear coupled optics. A universal and straightforward method to 
decouple two-dimensional harmonic oscillators with constant coefficients by symplectic 
transformations is presented, which is based on this isomorphism. The method yields
the eigenvalues (i.e. tunes) and eigenvectors and can be applied to a one-turn transfer matrix
or directly to the coefficient matrix of the linear differential equation.
\end{abstract}

\pacs{47.10.Df, 41.75.-i, 41.85.-p, 05.45.Xt, 03.30.+p, 03.65.Pm}
\keywords{Hamiltonian mechanics, particle beam focusing, coupled oscillators, Lorentz transformation, Dirac equation}
\maketitle

\section*{Introduction}

Even though there is continuous interest in this field (see, for instance,~\cite{Dattoli1,Dattoli2,Qin,Qin2,Qin3,Lebedev,Luo,Teng,EdwardsTeng,Brown0,Wolski}),
the treatment of coupled linear optics in two (or more) degrees of freedom has not 
yet reached the same level of generality, transparency and conceptual clarity as 
provided by the Courant-Synder theory for one degree of freedom.

This article is about coupled linear optics as required to describe, for instance, the 
motion of charged particles in accelerators and ion beam optics. 
Even though ion beam optics is in principle tree-dimensional, often symmetries can be 
used to reduce the problem to the treatment of two-dimensional systems.
In accelerators like cyclotrons or synchrotrons, the beam circulates in the 
horizontal plane and the electric and magnetic fields are symmetric with respect to 
this so-called {\it median plane}. In this case vertical motion is neither coupled 
to the horizontal nor to the longitudinal motion. But the dispersion of the bending 
magnets couples horizontal and longitudinal motion.
In other configurations, the transversal degrees of freedom are coupled with each 
other by solenoid magnets, but not with the longitudinal motion - or the longitudinal 
degree of freedom is ``hidden'' as the beam is not bunched but continuous. 
Therefore we will treat coupling in two dimensions like most authors~\cite{Dattoli1,Dattoli2,Qin,
Qin2,Qin3,Lebedev,Luo,Teng,EdwardsTeng,Brown0}. In an accompanying paper the
problem of transverse-longitudinal coupling by space charge forces in isochronous
cyclotrons is treated in linear approximation~\cite{cyc_paper}. This special type of 
coupling does not allow to apply the method of Teng and Edwards without
modifications~\cite{cyc_paper}. Other authors like Qin and Davidson restrict themselves 
to special forms of the Hamilton function so that their treatment lacks {\it generality}~\cite{Qin}.

The real Dirac matrices (RDMs) have been known for a long time as the {\it Majorana representation},
going back to a paper by Ettore Majorana~\cite{Majorana,Wilczek}. In this article,
we will use the RDMs in a very practical way that has - to the knowledge
of the author - not yet attracted much attention. The RDMs form a basis of 
$4\times 4$-matrices with remarkable properties also - and maybe especially -
in the context of classical mechanics. The RDMs are essential ingredients for
a formulation of a theory of symplectic flow in two dimensions. They enable
to survey all possible symplectic transformations in an elegant and straight manner.
 
The use of {\it real} instead of the complex Dirac matrices has several reasons: 
First, linear coupled optics is a classical theory and the relevant terms in the 
Hamilton function are real. Second, the RDMs are a {\it complete} basis: 
{\it Any} real $4\times 4$ matrix can be written as a linear combination of the RDMs. 
This in fact guarantees {\it generality} and {\it completeness} of the presented 
theory of linear coupled motion in two dimensions.
And finally - the RDMs are discriminable with respect to all relevant structural properties: 
Each matrix is either symplectic or antisymplectic, either even or odd, two RDMs 
either commute or anticommute. The use of the RDMs supports another clear distinction 
- real or imaginary. 

Furthermore the introduction of RDMs into classical mechanics 
may provide new insights into the relationship of Hamiltonian mechanics, special 
relativity and the Dirac equation.
It is known for a long time that the mathe\-matical formalism of Twiss-parameters and 
Courant-Snyder invariants can be applied to quantum systems~\cite{Dattoli3}. But 
no attempt has yet been made to apply the tools of quantum mechanics in classical 
mechanics.

Nevertheless we emphasize that a lot of the presented formalism, i.e. practically
all equations that do not refer to other matrices than $\y_0$, may be applied in 
arbitrary dimensions, if $\y_0$ is extended correspondingly. This holds especially 
for the concept of the {\it symplex}, that we introduce to identify the components 
of the ``force matrix''.
We will demonstrate the significance of the RDMs for the treatment of coupled
linear motion, in the context of transfer matrices and eigensystems. This leads 
in a natural way to the two-dimensional generalization of the Courant-Snyder
theory. In the second part we demonstrate how the RDMs can be used in the context
of the second moments using Poisson brackets. We show that both, the 
Maxwell-equations and the Lorentz force equations, can be formulated casually
in terms of the RDMs. This brings up an analogy which we call the ``electromechanical 
equivalence''. Effectively we use the isomorphism of symplectic transformations with 
Lorentz boosts and rotations in Minkowski space to introduce a physical nomenclature 
of symplectic transformations in two dimensions. Based on this equivalence and on 
the distinction of even and odd matrices we finally develope a general 
algorithm that allows to determine the symplectic decoupling transformation. 
Furthermore the algorithm enables to compute the eigenvalues and the eigenvectors 
of stable two-dimensional symplectic systems. 

\section{Real Dirac Matrices in Coupled Linear Optics}

The position and direction of a charged particle within a beam is usually described by 
its coordinates and angles relative to the reference trajectory.
In case of two transversal coordinates, the position of the particle is
described by $x,x',y,y'$, where $x$ is the horizontal and $y$ the vertical (``axial'')
coordinate. The dashes represent the derivative with respect to the path length $s$ 
along the reference orbit. In case of transverse-longitudinal coupling, a common choice
of the coordinates is $x,x',l,\delta$ where $l$ is the longitudinal coordinate with respect 
to the bunch center and $\delta={p-p_0\over p_0}$ is the relative momentum deviation,
where $p$ is the momentum of a specific ion and $p_0$ is the average momentum.

Since the formalism is related to classical Hamiltonian mechanics, we 
prefer to write $q_i$ and $p_i$ for the dynamical variables and
we collect them in a vector $\psi$: 
\begeq
\psi=(q_1,p_1,q_2,p_2)^T\,. 
\label{eq_stdvector}
\endeq
We could as well change the ordering of the variables and use for example
\begeq
\psi=(q_1,q_2,p_1,p_2)^T\,. 
\label{eq_alternatevector}
\endeq
This may (of course) have no influence on the physical situation and its description, 
but it leads to a different ordering of the RDMs. Four elements can be ordered in $4!=24$
different ways, but since we do not distinguish the indices, there are $4!/2=12$ 
different permutations of the elements in $\psi$ and corresponding systems of RDMs 
as listed in Tab.~\ref{tab_basis_sets}.

We identify the Dirac matrix $\y_0$ with the symplectic unit matrix, which
is usually denoted by $I$, $J$ or $S$.  
In case of Eq.~\ref{eq_stdvector}, the ``time direction'' $\y_0$ is
\begeq
\y_0=\bmtx{cccc}
0&1&0&0\\
-1&0&0&0\\
0&0&0&1\\
0&0&-1&0\\
\emtx\,,
\label{eq_gamma0_def}
\endeq
in case of Eq.~\ref{eq_alternatevector}, the form is
\begeq
\eta_0=\bmtx{cccc}
0&0&1&0\\
0&0&0&1\\
-1&0&0&0\\
0&-1&0&0\\
\emtx\,.
\endeq
$\y_0$ is a skew-symmetric matrix that squares to the negative unit matrix $\y_0^2=-{\bf 1}$,
while in the usual definition of the Dirac matrices one has $\y_0^2={\bf 1}$. The other three 
basic matrices $\y_1\dots\y_3$ are defined by the requirement that they must anticommute with 
$\y_0$ and with each other and that they square to the opposite sign of $\y_0^2$.
The signature of the metric tensor $g_{\mu\nu}$, hence, is $(-1,+1,+1,+1)$ - 
instead of $(+1,-1,-1,-1)$~\cite{Okubo,Scharnhorst}:
\begary{rcl}
g_{\mu\nu}&=&\mathrm{Diag}(-1,1,1,1)\\
          &=&{\y_\mu\,\y_\nu+\y_\nu\,\y_\mu\over 2}\,.
\label{eq_realmetric}
\endary
The other matrices that we use to form the symplectic basis are
\begeq
 \y_1=\bmtx{cccc}
   0 &  -1  &  0 &   0\\
  -1 &   0  &  0 &   0\\
   0 &   0  &  0 &   1\\
   0 &   0  &  1 &   0\\
\emtx
\endeq
\begeq
 \y_2=\bmtx{cccc}
   0 &   0  &  0 &   1\\
   0 &   0  &  1 &   0\\
   0 &   1  &  0 &   0\\
   1 &   0  &  0 &   0\\
\emtx
\endeq
\begeq
 \y_3=\bmtx{cccc}
  -1 &   0  &  0 &   0\\
   0 &   1  &  0 &   0\\
   0 &   0  & -1 &   0\\
   0 &   0  &  0 &   1\\
\emtx
\endeq
The remaining matrices are defined by
\begary{cccp{5mm}ccc}
\y_{14}&=&\y_0\,\y_1\,\y_2\,\y_3;&&\y_{15}&=&{\bf 1}\\
\y_4&=&\y_0\,\y_1;&&\y_7&=&\y_{14}\,\y_0\,\y_1=\y_2\,\y_3\\
\y_5&=&\y_0\,\y_2;&&\y_8&=&\y_{14}\,\y_0\,\y_2=\y_3\,\y_1\\
\y_6&=&\y_0\,\y_3;&&\y_9&=&\y_{14}\,\y_0\,\y_3=\y_1\,\y_2\\
\y_{10}&=&\y_{14}\,\y_0&=&\y_1\,\y_2\,\y_3&&\\
\y_{11}&=&\y_{14}\,\y_1&=&\y_0\,\y_2\,\y_3&&\\
\y_{12}&=&\y_{14}\,\y_2&=&\y_0\,\y_3\,\y_1&&\\
\y_{13}&=&\y_{14}\,\y_3&=&\y_0\,\y_1\,\y_2&&\\
\label{eq_gamma_def}
\endary
Note that the definition deviates from the conventions in 
particle physics, where the product $\y_0\,\y_1\,\y_2\,\y_3$
is usually labeled $\y_5$. The matrices are explicitly given in App.~\ref{sec_app1}.
If ${\bf A}$ is an arbitrary real-valued $4\times 4$ matrix, then ${\bf A}$ can be written as
a linear combination of RDMs
\begeq
{\bf A}=\sum\limits_{k=0}^{15}\,a_k\,\y_k\,,
\label{eq_analyse}
\endeq
where the RDM-coefficients $a_k$ are given by the {\it scalar product}
\begeq
a_k={\bf A}\,\cdot\,\y_k={(\y_k)^2\over 4}\,\mathrm{Tr}({{\bf A}\,\y_k+\y_k\,{\bf A}\over 2})\,,
\label{eq_scalarprod_def}
\endeq
and $\mathrm{Tr}({\bf X})$ is the trace of ${\bf X}$. 
The RDMs form a group and are the basis of a vector space. The associated algebra 
is the real Clifford algebra $Cl(3,1)$~\cite{Okubo,Scharnhorst}.
\begin{table}
\begin{tabular}{|l||l|c|c|c|c|c|c|c|c|c|c|}\hline
Type         &  $\y_x$                 & E.M.     & $\y_x^2$  & $\y_x^T$  & $\y_0$  & $\y_{14}$ & $\y_{10}$ & s/a & S/A & $\tilde\y_0$ & e/o\\\hline
$V_t$        &  $\y_0$                 &$\phi$    & $-1$      & $-$       & $-$     & $+$       & $+$       & s   & S   & y & e\\ 
$V_s$        &  $\vec\y$               &$\vec A$  & $+1$      & $+$       & $+$     & $+$       & $-$       & s   & A   & n & (e,o,e)\\ 
$B$          &  $\y_0\vec\y$           &$\vec E$  & $+1$      & $+$       & $+$     & $-$       & $+$       & s   & A   & n & (e,o,e)\\ 
$B$          &  $\y_{14}\y_0\vec\y$    &$\vec B$  & $-1$      & $-$       & $-$     & $-$       & $-$       & s   & S   & y & (o,e,o)\\ 
$A_t$        &  $\y_{14}\y_0$          &$\phi_m$  & $-1$      & $-$       & $+$     & $+$       & $-$       & a   & A   & y & o\\ 
$A_s$        &  $\y_{14}\vec\y$        &$\vec A_m$& $+1$      & $+$       & $-$     & $+$       & $+$       & a   & S   & n & (o,e,o)\\ 
$P$          &  $\y_{14}$              &$ $       & $-1$      & $-$       & $+$     & $-$       & $+$       & a   & A   & y & o\\ 
$S$          &  $\y_{15}={\bf 1}$      &$ $       & $+1$      & $+$       & $-$     & $-$       & $-$       & a   & S   & n & e\\ \hline
\end{tabular}
\caption{\label{tab_gamma} Properties of the real $\y$-matrices. The type encoding is $V$ for vectors, $B$ for bi-vector,
$A$ for axial vector, $P$ for pseudoscalar and $S$ for scalar and refers to the symplectic transformation properties. 
The subscripts $t$ ($s$) indicate time- (space-) like components of a four-vector, respectively. The column labeled $\y_a$ 
gives the sign $s$ that fulfills the following equation: $\y_a\,\y_x\,\y_a=s\,\y_x$. The column labeled ``s/a'' indicates, 
whether $\y_x$ is a symplex or an antisymplex, respectively. Accordingsly, the column labeled ``S/A'' tells, 
if $\y_x$ is symplectic or antisymplectic, respectively. The column labeled ``e/o'' indicates, whether the corresponding $\y$-matrix 
is even or odd. Even matrices are non-zero only in the block-diagonal components, odd matrices are zero in the block-diagonal 
components.
Finally, the column labeled  $\tilde\y_0$ denotes, if a basis exists such that $\y_x$ appears as the ``time direction'',
i.e. if a basis exists in which $\y_x$ plays the role of $\y_0$. This is only the case for antisymmetric matrices, 
which are either both a symplex {\it and} symplectic or none of it.}
\end{table}

\subsection{The Hamiltonian}

Linear coupled optics is characterized by a Hamiltonian function of the classical harmonic oscillator form 
\begeq
H={1\over 2}\,\psi^T\,{\bf A}\,\psi\,,
\label{eq_Hamiltonian}
\endeq
where the superscript ``T'' denotes a transposed vector or matrix. 
${\bf A}$ is a (generally time dependent) symmetric matrix.
The Hamilton equations of motion (EQOM) have the familiar form
\begary{rcl}
\dot q_i&=&{\d H\over \d p_i}\\
\dot p_i&=&-{\d H\over \d q_i}\,,
\label{eq_eqom_classical}
\endary
or, in vector notation,
\begeq
\dot\psi=\y_0\,\nabla_\psi\,H\,,
\label{eq_eqom_general}
\endeq
where the dot represents the time derivative.
It is well-known that the Jacobian Matrix ${\bf M}$ of a {\it canonical transformation} 
is symplectic, i.e. that~\cite{Arnold}
\begeq
{\bf M}\,\y_0\,{\bf M}^T=\y_0\,.
\label{eq_Msymplectic}
\endeq
A matrix ${\bf M}$ is antisymplectic, if
\begeq
{\bf M}\,\y_0\,{\bf M}^T=-\y_0\,.
\label{eq_Mantisymplectic}
\endeq
Eq.~\ref{eq_Msymplectic} includes also
\begary{rcl}
{\bf M}^T\,\y_0\,{\bf M}&=&\y_0\\
{\bf M}^T&=&-\y_0\,{\bf M}^{-1}\,\y_0\\
{\bf M}^{-1}&=&-\y_0\,{\bf M}^T\,\y_0\,.
\label{eq_Msymplectic2}
\endary

The product of symplectic matrices is symplectic. But note: Also the product 
of two anti-symplectic matrices is symplectic (see also Tab.~\ref{tab_gamma}).

\subsection{The Force Matrix and the Definition of a Symplex}
\label{sec_force_matrix}

From Eq.~\ref{eq_Hamiltonian} and Eq.~\ref{eq_eqom_general} one derives the following linear EQOM:
\begeq
\dot\psi=\y_0\,{\bf A}\,\psi={\bf F}\,\psi\,,
\label{eq_linear_eqom}
\endeq 
where ${\bf F}$ is the (generally time-dependent) force matrix.

The antisymmetry of $\y_0$ and the symmetry of ${\bf A}$ yield:
\begary{rcl}
{\bf F}^T&=&(\y_0\,{\bf A})^T\\
         &=&{\bf A}^T\,\y_0^T\\
         &=&-{\bf A}\,\y_0\\
         &=&\y_0\,\y_0\,{\bf A}\,\y_0\\
         &=&\y_0\,{\bf F}\,\y_0\\
\label{eq_symplex}
\endary
In the following we call a matrix ${\bf F}$ that fulfills EQ.~\ref{eq_symplex} a {\it symplex} 
(not ``simplex''). Symplices are sometimes called "infinitesimally symplectic" or "Hamiltonian"~\cite{Talman}, 
but the author prefers a unique and short name. A matrix ${\bf F}_a$ that holds 
\begary{rcl}
{\bf F}_a^T&=&-\y_0\,{\bf F}_a\,\y_0\\
\label{eq_antisymplex}
\endary
is called an {\it antisymplex}. $\y_0$ itself is a symplex as it is antisymmetric and squares to the 
negative unit matrix. By definition the basic matrices $\y_1\dots\y_3$ are also symplices.

The most important property of symplices is the {\it superposition principle}. 
According to this principle the sum of two symplices is a symplex: 
\begary{rcl}
({\bf F}_1+{\bf F}_2)^T&=&{\bf F}_1^T+{\bf F}_2^T=\y_0\,{\bf F}_1\,\y_0+\y_0\,{\bf F}_2\,\y_0\\
           &=&\y_0\,({\bf F}_1+{\bf F}_2)\,\y_0\,.
\label{eq_superposition}
\endary
The superposition principle includes scalability: A symplex multiplied by a scalar is still
a symplex. Given that the product of two symplices ${\bf F}_1$ and ${\bf F}_2$ is also a symplex, 
then one finds:
\begary{rcl}
\y_0\,({\bf F}_1\,{\bf F}_2)\,\y_0&=&({\bf F}_1\,{\bf F}_2)^T\\
                                  &=&{\bf F}_2^T\,{\bf F}_1^T\\
                                  &=&\y_0\,{\bf F}_2\,\y_0\,\y_0\,{\bf F}_1\,\y_0\\
                                  &=&-\y_0\,{\bf F}_2\,{\bf F}_1\,\y_0\\
{\bf F}_1\,{\bf F}_2&=&-{\bf F}_2\,{\bf F}_1\\
{\bf F}_1\,{\bf F}_2+{\bf F}_2\,{\bf F}_1&=&0\,.
\label{eq_symplex_ac}
\endary
The product of two symplices is a symplex, if (and only if) the symplices 
{\it anticommute}. Since the four basic matrices $\y_0\dots\y_3$ anticommute with each other, 
the six possible {\it bi-vectors} $\y_\nu\,\y_\mu$ are symplices. 

Since a symmetric $n\times n$-matrix ${\bf A}$ is described by $\nu$ parameters with
\begeq
\nu=n\,(n+1)/2\,,
\label{eq_dimensions}
\endeq
the force matrix ${\bf F}=\y_0\,{\bf A}$ must contain the same number of independent 
components. In case of $n=4$ we expect $\nu=10$ force components. These are
the four basic matrices, and the mentioned six {\it bi-vectors} 
$\y_\nu\,\y_\mu=\y_4\dots\y_9$. Hence, a general force matrix in two dimensions has the form
\begeq
{\bf F}=\sum\limits_{k=0}^{9}\,f_k\,\y_k\,.
\label{eq_gamma_comp}
\endeq

\subsection{Symmetric Products and Projectors}
\label{sec_projectors}

The symmetric product of a matrix ${\bf F}_1$, which is either a symplex or an 
antisymplex, and a symplex ${\bf F}_2$ is again a symplex:
\begary{rcl}
({\bf F}_1\,{\bf F}_2\,{\bf F}_1)^T&=&{\bf F}_1^T\,{\bf F}_2^T\,{\bf F}_1^T\\
&=&(\pm\,\y_0\,{\bf F}_1\,\y_0)\,(\y_0\,{\bf F}_2\,\y_0)\,(\pm \y_0\,{\bf F}_1\,\y_0)\\
&=&\y_0\,({\bf F}_1\,{\bf F}_2\,{\bf F}_1)\,\y_0\,.
\endary
Since all $\y$-matrices are either symplices or antisymplices, any expression of the form
$$
\sum\limits_{k=0}^{15}\,a_k\,\y_k\,{\bf F}\,\y_k
$$
with arbitrary coefficients $a_k$ is a symplex, if ${\bf F}$ is a symplex.
Tab.~\ref{tab_gamma} shows the result of $\y_a\y_x\y_a$ for $a=[0,10,14]$.
The result of the operation
\begeq
{\bf F}_a = {{\bf F} \pm \y_a\,{\bf F}\,\y_a\over 2}={1\over 2}\,\sum\limits_{k=0}^{9}\,f_k\,(\y_k\pm\y_a\,\y_k\,\y_a)\,,
\endeq
is a {\it projection}. For $a=14$ for instance one has
\begary{rcl}
{1\over 2}\,\sum\limits_{k=0}^{9}\,f_k\,(\y_k+\y_{14}\,\y_k\,\y_{14})&=&\sum\limits_{k=0}^{3}\,f_k\,\y_k\\
{1\over 2}\,\sum\limits_{k=0}^{9}\,f_k\,(\y_k-\y_{14}\,\y_k\,\y_{14})&=&\sum\limits_{k=4}^{9}\,f_k\,\y_k\,,
\endary
that is, $\y_{14}$ separates the ``vector''-components from the ``bi-vector'' components.

\subsection{The Transfer Matrix}
\label{sec_transfer_matrix}

The solution of Eq.~\ref{eq_linear_eqom} can be written by a symplectic {\it transfer matrix} 
${\bf M}(t,t_0)$:
\begeq
\psi(t)={\bf M}(t,t_0)\,\psi(t_0)\,.
\label{eq_transfer_matrix}
\endeq
If the force matrix is constant in time, then
\begeq
{\bf M}(t,t_0)=\exp{\left({\bf F}\,(t-t_0)\right)}\,.
\label{eq_transfer_matrix2}
\endeq
The time derivative of EQ.~\ref{eq_transfer_matrix} is:
\begary{rcl}
\dot\psi(t)&=&\dot{\bf M}(t,t_0)\,\psi(t_0)\\
           &=&{\bf F}\,\psi(t)\\
           &=&{\bf F}\,{\bf M}(t,t_0)\,\psi(t_0)\\
\endary
so that the following differential equation holds for ${\bf M}$:
\begeq
\dot{\bf M}={\bf F}\,{\bf M}\,.
\label{eq_transfer_matrix_deriv}
\endeq
Comparison with EQ.~\ref{eq_linear_eqom} shows that the column vectors of the
transfer matrix ${\bf M}$ are solutions of EQ.~\ref{eq_linear_eqom}.
If $n$ is the dimension of $\psi$, the complete transfer matrix can be 
obtained by integrating Eq.~\ref{eq_linear_eqom} $n$ times, 
using the $n$ euclidean unit vectors as starting conditions $\psi(t_0)$.

If the transfer matrix is expressed by a time-dependent matrix ${\bf\Phi}$ according to
\begary{rcl}
{\bf M}&=&\exp{{\bf\Phi}}=\sum\limits_{k=0}^\infty\,{{\bf\Phi}^k\over k!}\\
{\bf\dot M}&=&\dot{\bf\Phi}+{\dot{\bf\Phi}\,{\bf\Phi}+{\bf\Phi}\,\dot{\bf\Phi}\over 2}+{\dot{\bf\Phi}\,{\bf\Phi}^2+{\bf\Phi}\,\dot{\bf\Phi}\,{\bf\Phi}+{\bf\Phi}^2\,\dot{\bf\Phi}\over 6}+\dots\,,
\label{eq_dotM}
\endary
then if (and only if) ${\bf\Phi}$ and $\dot{\bf\Phi}$ commute, i.e. if 
\begeq
{\bf\Phi}\,\dot{\bf\Phi}=\dot{\bf\Phi}\,{\bf\Phi}\,,
\endeq 
Eq.~\ref{eq_dotM} can be written as:
\begary{rcl}
{\bf\dot M}&=&\dot{\bf\Phi}\,\left({\bf 1}+{\bf\Phi}+{{\bf\Phi}^2\over 2}+{{\bf\Phi}^3\over 6}+\dots\right)\\
           &=&\dot{\bf\Phi}\,\exp{{\bf\Phi}}\\
           &=&\dot{\bf\Phi}\,{\bf M}\\
\endary
so that in this case one finds
\begary{rcl}
{\bf F}&=&\dot{\bf\Phi}\\
{\bf\Phi}(t,t_0)&=&\int\limits_{t_0}^t\,{\bf F}(t)\,dt={\bf\bar F}\,\tau\,,
\endary
with $\tau=t-t_0$ so that the transfer matrix can be written as
\begeq
{\bf M}(t)=\exp{\left({\bf\bar F}\,\tau\,\right)}\,.
\label{eq_transfer_matrix_aver}
\endeq

If ${\bf\Phi}$ and $\dot{\bf\Phi}$ would {\it anticommute}, i.e. if
\begeq
{\bf\Phi}\,\dot{\bf\Phi}+\dot{\bf\Phi}\,{\bf\Phi}=0\,,
\label{eq_phdph_acomm}
\endeq 
then the square - and any even power of - ${\bf\Phi}$ would be constant and therefore one would find:
\begary{rcl}
{\bf\dot M}&=&\sum\limits_{k=0}^\infty\,{{\bf\Phi}^{2 k}\,{\bf\dot\Phi}\over (2k+1)!}=\sinh{(\bf\Phi)}\,{\bf\Phi}^{-1}\,{\bf\dot\Phi}\\
\endary
We define the matrices ${\bf M}_s$ and ${\bf M}_c$ according to
\begary{rcl}
{\bf M}&=&{\bf M}_s+{\bf M}_c\\
{\bf M}_s&\equiv&{\exp{({\bf\Phi})}-\exp{(-{\bf\Phi})}\over 2}={{\bf M}-{\bf M}^{-1}\over 2}=\sinh{(\bf\Phi)}\\
{\bf M}_c&\equiv&{\exp{({\bf\Phi})}+\exp{(-{\bf\Phi})}\over 2}={{\bf M}+{\bf M}^{-1}\over 2}=\cosh{(\bf\Phi)}\\
\label{eq_M_split}
\endary
so that Eq.~\ref{eq_phdph_acomm} would yield
\begary{rcl}
{\bf\dot M}_c&=&0\\
{\bf\dot M}_s&=&{\bf M}_s\,{\bf\Phi}^{-1}\,{\bf\dot\Phi}\,.
\label{eq_Mx1}
\endary
In Sec.~\ref{sec_M_forms} it will become clear that Eq.~\ref{eq_phdph_acomm} has to be rejected
in the case of stable focused systems.

Beam transfer lines and circular accelerators are typically composed of guiding ele\-ments that
provide a constant force matrix for a certain time (or better: length).
Examples are dipole- or quadrupole magnets, drifts and bends. In this case, it is possible to 
express the transfer matrix as a product of transfer-matrices for the individual elements:
\begary{rcl}
{\bf M}(t_n,t_0)&=&{\bf M}(t_n,t_{n-1})\dots{\bf M}(t_k,t_{k-1})\dots{\bf M}(t_1,t_0)\\
                &=&\exp{({\bf F}_n\,\tau_n)}\dots\exp{({\bf F}_1\,\tau_1)}\,.
\endary
The transfer matrix is symplectic, if ${\bf\Phi}$ is a symplex:
\begary{rcl}
{\bf M}\,\y_0\,{\bf M}^T &=&\left(\sum\limits_{k=0}^\infty\,{{\bf\Phi}^k\over k!}\right)\,\y_0\,\left(\sum\limits_{k=0}^\infty\,{({\bf\Phi}^T)^k\over k!}\right)\\
    &=&\left(\sum\limits_{k=0}^\infty\,{{\bf\Phi}^k\over k!}\right)\,\y_0\,\left(\sum\limits_{k=0}^\infty\,(-)^{k-1}\y_0\,{{\bf\Phi}^k\over k!}\,\y_0\,\right)\\
    &=&\left(\sum\limits_{k=0}^\infty\,{{\bf\Phi}^k\over k!}\right)\,\left(\sum\limits_{k=0}^\infty\,{(-{\bf\Phi})^k\over k!}\,\y_0\,\right)\\
    &=&\exp{\left({\bf\Phi}\right)}\,\exp{\left(-{\bf\Phi}\right)}\,\y_0\\
    &=&\y_0\\
\label{eq_M_symplectic}
\endary
{\it The exponential of a symplex is symplectic}.

\subsection{The Definition of Coupling}
\label{sec_coupling_def}

Before starting an investigation on decoupling there should be a clear definition
of {\it coupling}. A possible (and typical) definition refers to the 
structure of the force matrix. As we defined two degrees of freedom $q_1$ and $q_2$,
the obvious form of decoupling is a block-diagonal force matrix:
\begeq
{\bf F}=\bmtx{cc} {\bf A}& 0\\0&{\bf B}\emtx\,,
\endeq
where ${\bf A}$ and ${\bf B}$ are $2\times 2$-matrices.

There are less obvious forms of decoupling - consider a constant force matrix. 
The second derivative of the state vector $\psi$ is:
\begeq
\ddot\psi={\bf F}\,\dot\psi={\bf F}^2\,\psi\,.
\endeq
If the {\it square} of a constant force matrix is (block-) diagonal, then the system 
can be regarded as decoupled in second order. The solutions can be found separately 
for each degree of freedom. In this case the coupling fixes the relative
phases between the different degrees of freedom - but it does not determine
the functional form of the solution.

A force matrix of the Dirac-form is an example: 
\begary{rcl}
{\bf F}&=&E\,\y_0+p_1\,\y_1+p_2\,\y_2+p_3\,\y_3\\
{\bf F}^2&=&-(E^2-p_1^2-p_2^2-p_3^2)\,{\bf 1}\,.
\label{eq_dirac_op}
\endary
Even though the odd component $p_2\,\y_2$ is not block-diagonal,
the second order differential equation is decoupled. 
Another interesting example is a constant force matrix ${\bf X}$ of the form
\begary{rcl}
{\bf X}&=&-{\y_0-\y_2-\y_6-\y_7\over 2}\\
{\bf X}^2&=&-\y_{11}\\
{\bf X}^4&=&{\bf 1}\\
\endary
Here one finds that neither ${\bf X}$ nor ${\bf X}^2$ or ${\bf X}^3$ are block-diagonal.
Nevertheless the fourth time derivative of $\psi$ is ``decoupled''.

We refer to systems as {\it decoupled in n-th order}, if the n-th order EQOM have the form
\begeq
\left({d\over d\tau}\right)^n\,\psi={\bf B}\,\psi\,,
\endeq
where ${\bf B}$ is block-diagonal.
In case of time-dependent force matrices, the second derivative is
\begary{rcl}
\ddot\psi&=&{\bf\dot F}\,\psi+{\bf F}\,\dot\psi\\
         &=&({\bf\dot F}+{\bf F}^2)\,\psi\\
\endary
Hence we consider systems with time-dependent force matrices to be decoupled
to second order, if the expression ${\bf\dot F}+{\bf F}^2$ is block-diagonal.

\subsection{Symplectic Symplices}

Tab.~\ref{tab_gamma} lists all RDMs with their main properties. Each RDM ist either a 
symplex or an anti-symplex, symplectic or anti-symplectic. Four RDMs are both -
symplectic and symplex. If ${\bf F}$ is a symplectic symplex (SYSY), then the 
combination of EQ.~\ref{eq_M_symplectic} and EQ.~\ref{eq_symplex} yields:
\begary{rcl}
{\bf F}\,\y_0\,{\bf F}^T&=&{\bf F}\,\y_0\,\y_0\,{\bf F}\,\y_0=-{\bf F}^2\,\y_0=\y_0\\
\Rightarrow&&{\bf F}^2=-{\bf 1}\,.
\label{eq_sysy}
\endary
A symplex is symplectic, if (and only if) its square equals the negative unit matrix.
But note that symplectic matrices are {\it in principle} not scalable, i.e. without unit.
Symplices are scalable and may therefore have arbitrary units.
If the symplex ${\bf F}$ appears in the form of Eq.~\ref{eq_linear_eqom}, then it
has the natural unit of a frequency or wavenumber if the dot is interpreted as 
time or pathlength derivative, respectively. In practical problems of accelerator
physics, it will always be possible to find a typical length or time interval that
can be used to redefine the dot-derivative in such a way that the force matrix
is unitless. In consequence this means that force matrices which fulfill 
${\bf F}^2=-\W^2\,{\bf 1}$ with a constant $\W$ are equivalent to SYSYs:
\begary{rcl}
\dot\psi&=&{d\over dt}\psi={\bf F}\,\psi\\
\Rightarrow&&{d\over d\tau}\psi={{\bf F}\over\W}\,\psi\\
&&d\tau=\W\,dt\\
\endary
In this respect the Dirac-operator (Eq.~\ref{eq_dirac_op}) is a SYSY.

From Eq.~\ref{eq_sysy} and Eq.~\ref{eq_transfer_matrix2} it is quickly derived that the 
transfer matrix of a constant SYSY ${\bf F}$ is given by
\begary{rcl}
{\bf M}&=&{\bf 1}\,\cos{t}+{\bf F}\,\sin{t}\,.
\label{eq_cs_trans}
\endary
Eq.~\ref{eq_cs_trans} is known in the Courant-Synder theory of 1-dim. ion beam optics. 
The unitless parameter $t$ is called {\it phase advance}. If $t$ represents the phase advance
for one turn, then it is called {\it tune}. In the 1-dim. theory, the matrix ${\bf F}$
has the form
\begeq
{\bf F}=\bmtx{cc}
\alpha&\beta\\
-\y&-\alpha\\
\emtx\,,
\endeq
where $\alpha$, $\beta$ and $\y$ are the so-called Twiss-Parameters~\cite{Hinterberger}.
According to the concept of the Sec.~\ref{sec_coupling_def}, a constant SYSY is
decoupled in second order and Eq.~\ref{eq_cs_trans} shows that this definition is 
meaningful. Besides this a SYSY also has equal eigenfrequencies
- as will be shown in Sec.~\ref{sec_eigen} - and in consequence there is a common 
phase advance for both degrees of freedom.

Let ${\bf U}$ be a time dependent symplectic transformation, then
we obtain for the transformed ``spinor'' $\tilde\psi$:
\begary{rcl}
\tilde\psi&=&{\bf U}\,\psi\\
\dot{\tilde\psi}&=&\dot{\bf U}\,\psi+{\bf U}\,\dot\psi\\
\label{eq_sympltrafo}
\endary
The time derivative of EQ.~\ref{eq_linear_eqom} is
\begary{rcl}
\ddot\psi&=&\dot{\bf F}\,\psi+{\bf F}\,\dot\psi\\
\label{eq_eqom2}
\endary
The formal difference between Eq.~\ref{eq_sympltrafo} and Eq.~\ref{eq_eqom2}
is that ${\bf U}$ is symplectic and ${\bf F}$ is a symplex. If ${\bf U}$
is also a symplex (or if ${\bf F}$ is symplectic), then these equations
through light on the {\it structural equivalence} between a (time) derivative
and the multiplication with a SYSY. If the dynamics of a system is described
by a SYSY, then the (time) derivative is itsself a symplectic transformation. 
One could say that SYSYs are {\it operators}, which are equivalent 
to derivatives. 

\subsection{Eigenvalues and Eigenvectors}
\label{sec_eigen}

Eigenvalues and eigenvectors play an important role, if the system has some kind
of self-feedback. Circular accelerators like storage rings are a simple example 
for a system with self-feedback. Another example are systems with a constant 
force matrix, so-called ``constant focusing channels''.
If $\lambda$ is the diagonal matrix containing the eigenvalues of ${\bf F}$ and ${\bf E}$
is the matrix of columnwise eigenvectors, then 
\begeq
{\bf F}\,{\bf E}={\bf E}\,\lambda\,.
\endeq
If ${\bf E}$ can be reversed, then
\begeq
{\bf F}={\bf E}\,\lambda\,{\bf E}^{-1}\,.
\label{eq_F_eigen}
\endeq
In the simplest case of a constant force matrix one finds that
\begary{rcl}
{\bf M}(\tau,0)&=&\exp{\left({\bf E}\,\lambda\,{\bf E}^{-1}\,\tau\right)}\\
         &=&\sum\limits_{k=0}^\infty\,{({\bf E}\,\lambda\,{\bf E}^{-1}\,\tau)^k\over k!}\\
         &=&\sum\limits_{k=0}^\infty\,{\bf E}\,{\lambda^k\,\tau^k\over k!}\,{\bf E}^{-1}\\
         &=&{\bf E}\,\exp{(\lambda\,\tau)}\,{\bf E}^{-1}\\
         &=&{\bf E}\,\Lambda(\tau)\,{\bf E}^{-1}\,,
\label{eq_M_eigen}
\endary
where 
\begeq
\Lambda(\tau)=\exp{(\lambda\,\tau)}\,,
\label{eq_exp_eigenvalues}
\endeq 
is the diagonal matrix of the eigenvalues of ${\bf M}$.

The trace of the product of an antisymmetric and a symmetric matrix is zero. 
Hence ${\bf F}$ has zero trace. As similarity-transformations preserve the trace 
and the eigenvalues, we find  $Tr(\lambda)=Tr({\bf F})=0$ and hence the sum of all 
eigenvalues is zero. The eigenvalues are either real or (two) pairs of complex 
conjugate values. A symplectic transformation is said to be {\it strongly stable}, if
all eigenvalues of ${\bf M}$ are distinct and lie on the unit circle in the complex 
plane~\cite{Arnold}. This means that for stable (oscillatory) solutions the eigenvalues 
of ${\bf F}$ are two conjugate pairs of imaginary values:
\begary{rcl}
\lambda&=&\mathrm{Diag}(i\,\omega_1,-i\,\omega_1,i\,\omega_2,-i\,\omega_2)\\
       &=&-i\,{\omega_1+\omega_2\over 2}\,\y_3-i\,{\omega_1-\omega_2\over 2}\,\y_4\\
\label{eq_lambda}
\endary
Eq.~\ref{eq_F_eigen} yields:
\begary{rcl}
{\bf F}^2&=&{\bf E}\,\lambda^2\,{\bf E}^{-1}\\
         &=&-{\bf E}\,({\omega_1^2+\omega_2^2\over 2}\,{\bf 1}+{\omega_1^2-\omega_2^2\over 2}\,\y_{12})\,{\bf E}^{-1}\\
         &=&-{\omega_1^2+\omega_2^2\over 2}\,{\bf 1}-{\omega_1^2-\omega_2^2\over 2}\,{\bf E}\,\y_{12}\,{\bf E}^{-1}\,.
\endary
This shows that ${\bf F}$ is (isomorphic to) a SYSY, if $\omega_1^2=\omega_2^2$. 
SYSYs are degenerate. The absolute values of all eigenfrequencies of a SYSY are equal.

\subsection{The Form of the Transfer Matrix}
\label{sec_M_forms}

Eq.~\ref{eq_transfer_matrix_aver} allows specific functional forms $f(\tau)$ for
the matrix elements - depending on the dimensionality and the properties of the 
force matrix. The force matrix of a strongly stable system has eigenvalues that are 
grouped in two pairs of imaginary values - the eigenfrequencies. 
Ring-accelerators always have this property - but not in each section. 
Ion beam transport systems are usually composed of sections with separate 
elements like dipole-, quadrupole or sextupole magnets, drifts, solenoids, and so on. 
These elements are characterized by their individual transfer matrices and are 
not necessarily ``stable''. Only the product of the transfer matrices of all 
elements in a ring-accelerator has to be stable.
Eq.~\ref{eq_transfer_matrix_aver} is actually computed as a series:
\begeq
{\bf M}=\exp{({\bf\bar F}\,\tau)}=\sum\limits_{k=0}^\infty\,{({\bf\bar F}\,\tau)^k\over k!}\,.
\label{eq_M_expserie}
\endeq
In case of pure RDMs, i.e. ${\bf\bar F}=\y_k$ with $k\in [0\dots 9]$, 
the functional form of the elements of the corresponding transfer matrices 
are exponentials of $\tau$ - if we include (hyperbolic) sine- and cosine forms. 
In the case of arbitrary symplices, other forms are possible: 
A square matrix ${\bf F}$ is called {\it nilpotent}, if ${\bf F}^q=0$ for some
positive integer $q>1$. A simple example is the ``force'' matrix of a drift, which is
given by (see Eq.~\ref{eq_drift_mtx} in Sec.~\ref{sec_rdm_optics} below): 
\begary{rcl}
{\bf F}&=&{\y_0+\y_6\over 2}\\
{\bf M}&=&\sum\limits_{k=0}^\infty\,{({\bf\bar F}\,\tau)^k\over k!}\\
       &=&{\bf 1}+{\bf\bar F}\,\tau\,.
\endary
Since the maximal (non-zero) power of a $n\times n$ matrix ${\bf F}$ is $n-1$, 
polynomials up to third order as well as products of polynomials and exponentials 
are also possible solutions of Eq.~\ref{eq_M_expserie}.

In case of a constant force or in case of a transfer matrix for a complete {\it turn} in a stable circular 
accelerator, the diagonalized transfer matrix can be computed from Eq.~\ref{eq_exp_eigenvalues} and Eq.~\ref{eq_lambda}.
We introduce the abbreviations
\begary{rcl}
\Sigma_c&=&{\cos{(\omega_1\,\tau)}+\cos{(\omega_2\,\tau)}\over 2}=\cos{(\bar\omega\,\tau)}\,\cos{(\Delta\omega\,\tau)}\\
\Sigma_s&=&{\sin{(\omega_1\,\tau)}+\sin{(\omega_2\,\tau)}\over 2}=\sin{(\bar\omega\,\tau)}\,\cos{(\Delta\omega\,\tau)}\\
\Delta_s&=&{\sin{(\omega_1\,\tau)}-\sin{(\omega_2\,\tau)}\over 2}=\cos{(\bar\omega\,\tau)}\,\sin{(\Delta\omega\,\tau)}\\
\Delta_c&=&{\cos{(\omega_1\,\tau)}-\cos{(\omega_2\,\tau)}\over 2}=-\sin{(\bar\omega\,\tau)}\,\sin{(\Delta\omega\,\tau)}\,,
\label{eq_M_xcoeff}
\endary
where
\begary{rcl}
\bar\omega&=&{\omega_1+\omega_2\over 2}\\
\Delta\omega&=&{\omega_1-\omega_2\over 2}\\
\label{eq_M_xfreq}
\endary
so that the diagonal matrx $\Lambda$ (Eq.~\ref{eq_M_eigen} and Eq.~\ref{eq_exp_eigenvalues}) can be written as
\begeq
\Lambda=\Sigma_c\,{\bf 1}-i\,\Sigma_s\,\y_3-i\,\Delta_s\,\y_4-\Delta_c\,\y_{12}\,,
\endeq
and the one-turn-transfer matrix of strongly stable systems is according to Eq.~\ref{eq_M_eigen} given by:
\begary{rcl}
{\bf M}&=&\Sigma_c\,{\bf 1}-i\,\Sigma_s\,{\bf E}\,\y_3\,{\bf E}^{-1}\\
       &-&i\,\Delta_s\,{\bf E}\,\y_4\,{\bf E}^{-1}-\Delta_c\,{\bf E}\,\y_{12}\,{\bf E}^{-1}\\
\label{eq_M_xeigen}
\endary
Eq.~\ref{eq_M_xeigen} is the generalization of the Twiss-matrix for 2-dimensional systems~\cite{CS}. 
It shows that transfer matrices usually have - in contrast to symplices - a scalar component as well as
components which are antisymplices.
The second and the third term of Eq.~\ref{eq_M_xeigen} have the
same {\it form} as a force matrix (though different eigenvalues). The last term
vanishes in case of a degenerate system with equal eigenfrequencies. In this case 
the one-turn transfer matrix has the form of Eq.~\ref{eq_cs_trans}. 

In order to split the transfer matrix into the components according to Eq.~\ref{eq_M_xeigen},
we make use of the method of projections, modified for this purpose, as follows:
The inverse of a symplectic (transfer-) matrix ${\bf M}$ has the same eigenvectors as ${\bf M}$:
\begary{rcl}
{\bf M}^{-1}&=&{\bf E}\,\Lambda^{-1}(\tau)\,{\bf E}^{-1}\\
            &=&{\bf E}\,\Lambda(-\tau)\,{\bf E}^{-1}={\bf E}\,\exp{(-\lambda\,\tau)}\,{\bf E}^{-1}\,,
\endary
so that using Eq.~\ref{eq_Msymplectic2} yields:
\begary{rcl}
{1\over 2}\,({\bf M}\pm{\bf M}^{-1})&=&{\bf E}\,{\Lambda(\tau)\pm\Lambda(-\tau)\over 2}\,{\bf E}^{-1}\\
                             &=&{1\over 2}\,({\bf M}\mp\y_0\,{\bf M}^T\,\y_0)\,.
\label{eq_M_filter}
\endary
If we recall Eq.~\ref{eq_M_split}, then we can determine the RDM-coefficients of the transfer 
matrix and the matrices ${\bf M}_s$ and ${\bf M}_c$ according to Eq.~\ref{eq_analyse} and 
Eq.~\ref{eq_scalarprod_def}. Let $m_k$ be the resulting coefficients, then 
insertion into Eq.~\ref{eq_M_filter} yields:
\begary{rcl}
{\bf M}_s&=&{1\over 2}\,({\bf M}+\y_0\,{\bf M}^T\,\y_0)=\sum\limits_{k=0}^{9}\,m_k\,\y_k\\
                                           &=&{\bf E}\,{\Lambda(\tau)-\Lambda(-\tau)\over 2}\,{\bf E}^{-1}\\
                                           &=&-i\,\Sigma_s\,{\bf E}\,\y_3\,{\bf E}^{-1}-i\,\Delta_s\,{\bf E}\,\y_4\,{\bf E}^{-1}\\
{\bf M}_c&=&{1\over 2}\,({\bf M}-\y_0\,{\bf M}^T\,\y_0)=\sum\limits_{k=10}^{15}\,m_k\,\y_k\\
                                           &=&{\bf E}\,{\Lambda(\tau)+\Lambda(-\tau)\over 2}\,{\bf E}^{-1}\\
                                           &=&\Sigma_c\,{\bf 1}-\Delta_c\,{\bf E}\,\y_{12}\,{\bf E}^{-1}\\
\label{eq_M_filter2}
\endary
A decoupled force matrix logically implies a decoupled transfer matrix. 
A comparison of Eq.~\ref{eq_F_eigen}, Eq.~\ref{eq_lambda} and Eq.~\ref{eq_M_filter2} shows
that the force matrix and ${\bf M}_s$ differ only in the eigenvalues. Hence all the information 
that is required to compute the decoupling transformation can be obtained either from the force matrix
or from ${\bf M}_s$. ${\bf M}_c$ can be ignored in the context of decoupling. Even more than that:
The matrix of eigenvectors ${\bf E}$ of the matrix ${\bf M}_s$ diagonalizes ${\bf M}_c$, too. 
We come back to this in Sec.~\ref{sec_diag} and Sec.~\ref{sec_M_compounds}, after the construction 
of the matrix of eigenvectors ${\bf E}$.

From Eq.~\ref{eq_M_filter2} one quickly derives that ${\bf M}_s$ is a symplex while ${\bf M}_c$ is an antisymplex.
As ${\bf M}_s$ and ${\bf M}_c$ share the same eigenvectors and since all diagonal matrices, 
i.e. $\y_3$, $\y_4$, $\y_{12}$ and the unit matrix commute, also ${\bf M}_s$ and ${\bf M}_c$ commute:
\begeq
{\bf M}_s\,{\bf M}_c-{\bf M}_c\,{\bf M}_s=0\,.
\label{eq_MsMcComm}
\endeq
From Eq.~\ref{eq_Msymplectic} and Eq.~\ref{eq_MsMcComm} one derives
\begeq
{\bf M}_c^2-{\bf M}_s^2={\bf 1}\,,
\endeq
which includes that
\begary{rcl}
1&=&\Sigma_c^2+\Delta_c^2+\Sigma_s^2+\Delta_s^2\\
0&=&\Sigma_c\,\Delta_c+\Sigma_s\,\Delta_s\,,
\endary
in agreement with Eq.~\ref{eq_M_xcoeff}.

\subsection{Second Moments and the Envelope Equations}

Besides the position of a beam relative to the design orbit, the most important
properties of an ion beam are collected in the matrix of second moments.
We assume in the following that the beam is centered, i.e. that the first moments
are all identically zero. 
If the state vectors $\psi_i(\tau)$ with $i=1\dots n$ represent a family of $n$ ions,
where $\tau$ is the pathlength along the reference orbit, then the matrix of
second moments ${\bf\sigma}$ is given by
\begeq
{\bf\sigma}={1\over n}\,\sum\limits_{i=1}^n\,\psi_i\,\psi^T_i=\langle \psi\,\psi^T\rangle\,.
\label{eq_sigma_def}
\endeq
Another possible parameterization is given by a density function $\rho(\psi,\tau)=\rho(q_1,p_1,q_2,p_2,\tau)$
which should be normalized such that
\begeq
\int\dots\int\,\rho(q_1,p_1,q_2,p_2,\tau)\,dq_1\,dp_1\,dq_2\,dp_2=1\,.
\endeq
In this case one writes
\begeq
{\bf\sigma}=\int\dots\int\,\rho\,\psi\,\psi^T\,dq_1\,dp_1\,dq_2\,dp_2\,.
\label{eq_sigma_rho_int}
\endeq
But independent of the specific practical method of computation, we assume that 
the matrix of second moments $\sigma=\langle\,\psi\,\psi^T\,\rangle$ is well-defined
and has a non-vanishing determinant.
From Eq.~\ref{eq_linear_eqom} one derives:
\begary{rcl}
\dot\sigma&=&\langle\,\dot\psi\,\psi^T\,\rangle+\langle\,\psi\,\dot\psi^T\,\rangle\\
          &=&\langle\,{\bf F}\,\psi\,\psi^T\,\rangle+\langle\,\psi\,\psi^T\,{\bf F}^T\,\rangle\\
          &=&{\bf F}\,\sigma+\sigma\,{\bf F}^T\\
          &=&{\bf F}\,\sigma+\sigma\,\y_0\,{\bf F}\,\y_0\\
\label{eq_dotsigma}
\endary
We define the {\bf S}-matrix using $\bar\psi\equiv\psi^T\,\y_0$ by
\begeq
{\bf S}\equiv\sigma\,\y_0=\langle\,\psi\,\psi^T\,\y_0\,\rangle=\langle\,\psi\,\bar\psi\,\rangle\,,
\endeq
and obtain from Eq.~\ref{eq_dotsigma} by multiplication from the right with $\y_0$:
\begary{rcl}
{\bf\dot S}&=&{\bf F}\,{\bf S}-{\bf S}\,{\bf F}\\
\label{eq_dotS}
\endary
In ion beam physics Eq.~\ref{eq_dotsigma} is called an {\it envelope equation}, as
the second moments define the envelope of an ion beam. 
${\bf S}$ is a symplex - as any symmetric matrix multiplied by $\y_0$:
\begary{rcl}
{\bf S}^T&=&\y_0^T\,\sigma^T\\
         &=&-\y_0\,\sigma\\
         &=&\y_0\,\sigma\,\y_0^2\\
         &=&\y_0\,{\bf S}\,\y_0\,.
\endary
Assuming a constant force matrix ${\bf F}$, Eq.~\ref{eq_dotS} tells us that the 
second moments are constant, if ${\bf S}$ and ${\bf F}$ commute. Using the
eigenvector-decomposition Eq.~\ref{eq_F_eigen} of ${\bf F}$ gives
\begary{rcl}
0&=&{\bf F}\,{\bf S}-{\bf S}\,{\bf F}\\
 &=&{\bf E}\,\lambda\,{\bf E}^{-1}\,{\bf S}-{\bf S}\,{\bf E}\,\lambda\,{\bf E}^{-1}\\
0&=&\lambda\,{\bf E}^{-1}\,{\bf S}\,{\bf E}-{\bf E}^{-1}\,{\bf S}\,{\bf E}\,\lambda\\
0&=&\lambda\,{\bf \tilde S}-{\bf\tilde S}\,\lambda\\
\label{eq_eigen1}
\endary
Since $\lambda$ is diagonal and ${\bf\tilde S}$ commutes with $\lambda$, also 
\begeq
{\bf\tilde S}={\bf E}^{-1}\,{\bf S}\,{\bf E}={\bf D}\,,
\endeq 
must be diagonal (see Tab.~\ref{tab_commtab}), so that the matrix ${\bf S}$ has the form:
\begeq
{\bf S}={\bf E}\,{\bf D}\,{\bf E}^{-1}\,.
\label{eq_S_eigen}
\endeq 
The force matrix and the ${\bf S}$-matrix share the same eigenvectors -
but will in general have different eigenvalues. 

\subsection{Matching}

Matching is a concept of circular accelerators, i.e. systems with self-feedback, where eigenvalues 
and -vectors are well defined.
A beam is matched, if the phase space occupied by the ions in the beam fits to the ``acceptance'' of the 
machine. The practical consequence of mismatching is an oscillation or ``pumping'' of the phase space 
distribution which typically leads to an increase of the beam emittance by filamentation~\cite{Hinterberger}. 

Wolski formulated the state of matching for the general case of periodic motion~\cite{Wolski}.
In this case, the restriction is no more that ${\bf F}$ has to be constant, but that ${\bf F}$
must be periodic: ${\bf F}(\tau+C)={\bf F}(\tau)$ for a given period $C$ and any $\tau$.
Given the transfer matrix over one turn (or period) is ${\bf M}$, then the beam is matched, if
\begeq
\sigma={\bf M}\,\sigma\,{\bf M}^T\,.
\endeq
Using Eq.~\ref{eq_Msymplectic2}, one quickly finds in analogy to Eq.~\ref{eq_eigen1}:
\begary{rcl}
\sigma&=&-{\bf M}\,\sigma\,\y_0\,{\bf M}^{-1}\,\y_0\\
\sigma\,\y_0&=&{\bf M}\,\sigma\,\y_0\,{\bf M}^{-1}\\
{\bf S}&=&{\bf M}\,{\bf S}\,{\bf M}^{-1}\\
{\bf S}\,{\bf M}&=&{\bf M}\,{\bf S}\\
0&=&{\bf M}\,{\bf S}-{\bf S}\,{\bf M}\\
\endary
Following the same arguments as for Eq.~\ref{eq_eigen1} one finds that the matrix of 
second moments is {\it matched}, if the transfer matrix over one period 
and the ${\bf S}$-matrix share the same system of eigenvectors.
The general form of the matrix ${\bf S}$ is again given by Eq.~\ref{eq_S_eigen}
where ${\bf D}$ has a form analogue to Eq.~\ref{eq_lambda} and is given by~\cite{Wolski}:
\begeq
{\bf D}=\mathrm{Diag}(i\,\eps_1,-i\,\eps_1,i\,\eps_2,-i\,\eps_2)\,,
\label{eq_Sdiag}
\endeq
where the $\eps_i$ are the emittances of the two degrees of freedom.

\subsection{Expectation Values and Scalar Product}

The ``expectation values'' $\langle\bar\psi\,\y_x\,\psi\rangle$ are related 
to the ${\bf S}$-matrix according to:
\begary{rcl}
\bar\psi\,\y_x\,\psi&=&\sum\limits_{ijk}\,\psi^k\,\y_0^{ki}\,\y_x^{ij}\,\psi^j\\
      &=&{1\over 2}\,\sum\limits_{ijk}\,(\psi^k\,\y_0^{ki}\,\y_x^{ij}\,\psi^j+\psi^j\,\y_0^{jk}\,\y_x^{ki}\,\psi^i)\\
      &=&{1\over 2}\,\sum\limits_{ijk}\,(\y_x^{ij}\,\psi^j\,\psi^k\,\y_0^{ki}+\psi^i\,\psi^j\,\y_0^{jk}\,\y_x^{ki})\\
      &=&{1\over 2}\,Tr\left(\sum\limits_{jk}\,(\y_x^{ij}\,\psi^j\,\psi^k\,\y_0^{kl}+\psi^i\,\psi^j\,\y_0^{jk}\,\y_x^{kl})\right)\\
      &=&{1\over 2}\,Tr(\y_x\,\psi\,\bar\psi+\psi\,\bar\psi\,\y_x)\\
      &=&{1\over 2}\,Tr(\y_x\,{\bf S}+{\bf S}\,\y_x)\,.
\label{eq_scalar_vs_ev}
\endary
The RDM-coefficients of the ${\bf S}$-matrix are - apart form the sign -
the expectation values of the RDMs (see Eq.~\ref{eq_scalarprod_def}).

\section{Poisson Brackets of Second Moments and the Electromechanical Equivalence}
\label{sec_PoissionBrackets}

The total time derivative of a function $f(p,q,t)$ is given by the Poisson-brackets with the
Hamiltonian function~\cite{Goldstein}:
\begary{rcl}
{df(q_i,p_i,t)\over dt}&=&{\d f\over \d t}+\sum\limits_i\left\{{\d f\over \d q_i}{\d H\over \d p_i}-{\d f\over \d p_i}{\d H\over \d q_i}\right\}\\
                     &=&
\bmtx{c}
{\d f\over \d q_1}\\
{\d f\over \d p_1}\\
{\d f\over \d q_2}\\
{\d f\over \d p_2}\\
\emtx\,\y_0\,
\bmtx{c}
{\d H\over \d q_1}\\
{\d H\over \d p_1}\\
{\d H\over \d q_2}\\
{\d H\over \d p_2}\\
\emtx\\
                     &=&{\d f\over \d t}+\nabla_{q,p}\,f(p,q,t)\,\dot\psi\\
                     &=&{\d f\over \d t}+\nabla_{q,p}\,f(p,q,t)\,{\bf F}\,\psi\,.
\label{eq_poisson_brackets}
\endary
We define the functions $f_k$ by the ``expectation values'' according to
\begeq
f_k(p,q)={1\over 2}\,\bar\psi\,\y_k\,\psi\,.
\endeq
$f_k(p,q)$ do not explicitely depend on time. Evidently the $f_k$ vanish for all
non-symmetric matrices $\y_0\,\y_k$ so that there should be exactly $n\,(n+1)/2=10$ 
non-vanishing functions $f_k$. The gradient $\nabla_{p,q}=\nabla_\psi$ yields
\begeq
\d_{\psi_i}\psi_j=\delta_{ij}\,,
\endeq
so that 
\begary{rcl}
\nabla_{p,q}\,f_k&=&{1\over 2}\,\left(\y_0\,\y_k\,\psi+\psi^T\,\y_0\,\y_k\right)\\
                 &=&{1\over 2}\,\psi^T\,\left(\y_0\,\y_k+(\y_0\,\y_k)^T\right)\\
                 &=&{1\over 2}\,\psi^T\,\left(\y_0\,\y_k+\y_k^T\,\y_0^T\right)\\
                 &=&{1\over 2}\,\psi^T\,\left(\y_0\,\y_k-\y_k^T\,\y_0\right)\\
                 &=&{1\over 2}\,\psi^T\,\left(\y_0\,\y_k+\y_0\,\y_0\y_k^T\,\y_0\right)\\
                 &=&{1\over 2}\,\bar\psi\,\left(\y_k+\y_0\y_k^T\,\y_0\right)\,.
\endary
Since all $\y$-matrices are either symplices or antisymplices, it is obvious, that
\begeq
\nabla_{p,q}\,f_k=\left\{\begin{array}{rcl}
\bar\psi\,\y_k&\mathrm{for}&k\in [0\dots 9]\\
             0&\mathrm{for}&k\in [10\dots 15]\\
\end{array}\right.\,,
\endeq
i.e. {\it only symplices} can have a non-vanishing expectation value.
For all symplices $\y_k$ with $k\in\,[0\dots 9]$ the Poisson brackets result:
\begeq
{d\over d\tau}\,\left({\bar\psi\y_k\,\psi\over 2}\right)=\bar\psi\,\y_k\,{\bf F}\,\psi\,.
\label{eq_dt_poisson}
\endeq
On the other hand we find from EQ~\ref{eq_linear_eqom}:
\begary{rcl}
{d\over d\tau}\,\left(\bar\psi\y_k\,\psi\right)&=&\dot\psi^T\,\y_0\,\y_k\,\psi+\psi^T\,\y_0\,\y_k\,\dot\psi\\
          &=&\psi^T\,{\bf F}^T\,\y_0\,\y_k\,\psi+\psi^T\,\y_0\,\y_k\,{\bf F}\,\psi\\
          &=&\psi^T\,\y_0\,{\bf F}\,\y_0\,\y_0\,\y_k\,\psi+\psi^T\,\y_0\,\y_k\,{\bf F}\,\psi\\
          &=&-\bar\psi\,{\bf F}\,\y_k\,\psi+\bar\psi\,\y_k\,{\bf F}\,\psi\\
          &=&\bar\psi\,(\y_k\,{\bf F}-{\bf F}\,\y_k)\,\psi\\
\label{eq_dt_commutator}
\endary
Combining EQ.~\ref{eq_dt_poisson} and~\ref{eq_dt_commutator} yields:
\begeq
{d\over d\tau}\,\left(\bar\psi\y_k\,\psi\right)=\bar\psi\,2\,\y_k\,{\bf F}\,\psi=\bar\psi\,(\y_k\,{\bf F}-{\bf F}\,\y_k)\,\psi\,,
\endeq
so that
\begary{rcl}
0&=&\bar\psi\,2\,\y_k\,{\bf F}\,\psi-\bar\psi\,(\y_k\,{\bf F}-{\bf F}\,\y_k)\,\psi\\
0&=&\bar\psi\,(\y_k\,{\bf F}+{\bf F}\,\y_k)\,\psi\\
\label{eq_symm_death}
\endary

According to EQ.~\ref{eq_gamma_comp}, the force matrix ${\bf F}$ can be written as
\begeq
{\bf F}=\sum\limits_{l=0}^{9}\,F_l\,\y_l\,,
\endeq 
so that EQ.~\ref{eq_dt_commutator} can be written as
\begary{rcl}
\dot f_k&=&{d\over d\tau}\,\left({\bar\psi\y_k\,\psi\over 2}\right)=\sum\limits_{l=0}^{9}\,\bar\psi\,{\y_k\,\y_l-\y_l\,\y_k\over 2}\,\psi\,F_l\\
&=&\sum\limits_{l=0}^{9}\,{\bf G}_{kl}\,F_l\,.
\label{eq_fk_Gkl_Fl}
\endary
The matrix ${\bf G}_{kl}$ is composed of the expectation values $f_i$
and the indices of $f_i$ are given by the upper left $10\times 10$ part of 
the commutator table (Tab.~\ref{tab_commtab}).
Note that all commutators of symplices are either zero or again symplices.

Eq.~\ref{eq_fk_Gkl_Fl} can be reordered such that it has the form of a linear 
transformation of a 10-dimensional vector $f_i$:
\begeq
\dot f_i=\sum\limits_{j=0}^9\,{\bf B}_{ij}\,f_j\,.
\endeq
The matrix ${\bf G}_{kl}$ is antisymmetric. The reordering results
in a quite similar matrix ${\bf B}_{ij}$.
Now we introduce a normalization that assures a positive sign for
positive definite second moments. For instance $f_0$ is according to
the definition given by
\begeq
f_0=\bar\psi\,\y_0\,\psi=\psi^T\,\y_0^2\,\psi=-\psi^T\,\psi\,,
\endeq
which is negative even though the corresponding second moment
is positive. The normalization is done by multiplication with
$\y_k^2=\pm\,1$, which equals $-1$ for $k\,\in\,\{0,7,8,9\}$.
This can be expressed by the multiplication with a quadratic 
diagonal matrix $\tilde g$ that is an extended version
of the ``metric tensor'' $g_{\mu\nu}$:
\begary{rcl}
\tilde g&=&\mathrm{Diag}(-1,1,1,1,1,1,1,-1,-1,-1)\\
\tilde g^2&=&{\bf 1}\\
\endary
The transformed EQOM are:
\begary{rcl}
\dot f&=&{\bf B}\,f\\
\tilde g\,\dot f&=&(\tilde g\,{\bf B}\,\tilde g)\,(\tilde g\,f)\\
\dot{\tilde f}&=&{\bf\tilde B}\,\tilde f\,,
\label{eq_eqom_2nd}
\endary
where the matrix ${\bf\tilde B}$ is explicitely given by 
{\tiny\begary{rcl}
{\bf\tilde B}&=&
\bmtx{cccccccccc}
     &  F_4&  F_5&  F_6& -F_1& -F_2& -F_3&     &     &      \\
  F_4&     &  F_9& -F_8& -F_0 &     &     &     &  F_3& -F_2 \\
  F_5& -F_9&     &  F_7&     & -F_0 &     & -F_3&     &  F_1 \\
  F_6&  F_8& -F_7&     &     &     & -F_0 &  F_2& -F_1&      \\
 -F_1&  F_0&     &     &     &  F_9& -F_8&     &  F_6& -F_5 \\
 -F_2&     &  F_0&     & -F_9&     &  F_7& -F_6&     &  F_4 \\
 -F_3&     &     &  F_0&  F_8& -F_7&     &  F_5& -F_4&      \\
     &     & -F_3&  F_2&     & -F_6&  F_5&     &  F_9& -F_8 \\
     &  F_3&     & -F_1&  F_6&     & -F_4& -F_9&     &  F_7 \\
     & -F_2&  F_1&     & -F_5&  F_4&     &  F_8& -F_7&      \\
\emtx\\
\label{eq_envelope_forces}
\endary}
Eq.~\ref{eq_eqom_2nd} is another way to express the envelope equations
(Eq.~\ref{eq_dotsigma} and Eq.~\ref{eq_dotS}). The explicite relation
between the second moments $\sigma_{ij}$ and the expectation values
is given in App.~\ref{sec_app4}.

\subsection{Symplectic Electrodynamics}
\label{sec_ED}

The upper left $4\times 4$-block of the matrix ${\bf\tilde B}$ equals the 
electromagnetic field tensor, if we replace the ``vector'' $(F_4,F_5,F_6)^T$ 
by the electric field $\vec E$  and  $(F_7,F_8,F_9)^T$ by the magnetic field $\vec B$.
In this section we investigate the equivalence of two-dimensional symplectic flow 
and relativistic electrodynamics. In the next section we show that rotations
and Lorentz boosts in Minkowski space are formally identical to a subset of symplectic 
transformations in two-dimensional coupled linear optics.

If one writes the 4-potential ${\bf\Phi}$ (4-current ${\bf J}$, 4-momentum ${\bf P}$)
as a 4-vector using the RDMs $\y_0\dots\y_3$ according to
\begeq
{\bf\Phi}=\phi\,\y_0+A_x\,\y_1+A_y\,\y_2+A_z\,\y_3=\phi\,\y_0+\vec\y\,\vec A\,,
\endeq
the 4-derivative ${\bf D}$ as
\begeq
{\bf D}=\d_t\,\y_0-\d_x\,\y_1-\d_y\,\y_2-\d_z\,\y_3\,,
\endeq
and the electromagnetic fields $\vec E$ and $\vec B$ as
\begeq
{\bf F}=E_x\,\y_4+E_y\,\y_5+E_z\,\y_6+B_x\,\y_7+B_y\,\y_8+B_z\,\y_9\,,
\endeq
then the Maxwell equations (MWEQs) can be written (remarkably compact) as:
\begary{rcl}
{\bf F}&=&-{\bf D}\,{\bf\Phi}\\
{\bf D}\,{\bf F}&=&4\,\pi\,{\bf J}\,,
\label{eq_maxwell}
\endary
with the usual choice of units. 

The Lorentz force can also be expressed by RDMs. If the 4-momentum is defined by
\begeq
{\bf P}={\cal E}\,\y_0+p_x\,\y_1+p_y\,\y_2+p_z\,\y_3={\cal E}\,\y_0+\vec p\,\vec\y\,,
\endeq
- where ${\cal E}$ is the energy and $\vec p$ the momentum -
then the Lorentz force equations can be written as
\begeq
{d{\bf P}\over d\tau}=\dot{\bf P}={q\over 2\,m}\left({\bf F}\,{\bf P}-{\bf P}\,{\bf F}\right)\,,
\label{eq_lorentzforce}
\endeq
where $\tau$ is the proper time.
In the lab frame time $dt={d\tau\over\y}$ EQ.~\ref{eq_lorentzforce} yields (setting $c=1$):
\begary{rcl}
{d{\cal E}\over d\tau}&=&{q\over m}\,\vec p\,\vec E\\
{d\vec p\over d\tau}&=&{q\over m}\,\left({\cal E}\,\vec E+\vec p\times\vec B\right)\\
\y\,{d{\cal E}\over dt}&=&q\,\y\,\vec v\,\vec E\\
\y\,{d\vec p\over dt}&=&{q\over m}\,\left(m\,\y\,\vec E+m\,\y\,\vec v\times\vec B\right)\\
{d E\over dt}&=&q\,\vec v\,\vec E\\
{d\vec p\over dt}&=&q\,\left(\vec E+\vec v\times\vec B\right)\,,
\endary
which are exactly the Lorentz force equations.

We use the formal identity of Eq.~\ref{eq_dotS} and Eq.~\ref{eq_lorentzforce} to
introduce the {\bf electromechanical equivalence} (EMEQ) of 2-dim. symplectic motion
and the motion of charged particles in electromagnetic fields as described by
the Lorentz force equations. The statement of the EMEQ is, that the transformation 
properties of the elements of the ${\bf S}$-matrix with respect to symplectic transformations
are {\it formally} identical to the transformation properties of momentum ${\bf P}$ and 
electromagnetic field ${\bf F}$ in Minkowski space. The analogy allows to obtain an overview 
over the components of the force matrix ${\bf F}$ and to establish a meaningful 
nomenclature for the RDM-coefficients: The ten elements of the force matrix are 
associated with the energy and momentum of - and with 
electric and magnetic field components ``seen by'' - a relativistic charged particle:
\begary{rcl}
{\bf F}&=&{\cal E}\,\y_0+p_x\,\y_1+p_y\,\y_2+p_z\,\y_3\\
       &+&E_x\,\y_4+E_y\,\y_5+E_z\,\y_6\\
       &+&B_x\,\y_7+B_y\,\y_8+B_z\,\y_9\,.
\label{eq_edeq}
\endary
If the force matrix is split into the electrodynamical ${\bf F}_{ed}$ and the 
mechanical components ${\bf F}_m$, the matrices are explicitely given by 
{\small\begary{rcl}
{\bf F}&=&{\bf F}_{ed}+{\bf F}_m\\
&=&\bmtx{cccc}
-E_x&E_z+B_y&E_y-B_z&B_x\\
E_z-B_y&E_x&-B_x&-E_y-B_z\\
E_y+B_z&B_x&E_x&E_z-B_y\\
-B_x&-E_y+B_z&E_z+B_y&-E_x\\
\emtx\\
&+&\bmtx{cccc}
-p_z&{\cal E}-p_x&0&p_y\\
-{\cal E}-p_x&p_z&-p_y&0\\
0&p_y&-p_z&{\cal E}+p_x\\
p_y&0&-{\cal E}+p_x&p_z\\
\emtx\,.
\label{eq_edeq1}
\endary}
Note that there are ten force terms in symplectic motion and only six electromagnetic field components.
Correspondingly there are ten symplectic transformations in 2-dim. symplectic flow, but only
six transformations known in Minkowski space - rotations about and boosts along three axis.
One also finds that Eq.~\ref{eq_eqom_2nd} and Eq.~\ref{eq_envelope_forces} correspond to the
Lorentz force equations only, if $F0$, $F1$, $F2$ and $F3$ are zero. In this case ${\bf\tilde B}$ 
is block-diagonal - the first block being the upper left $4\times 4$-matrix and the
second block the lower right $6\times 6$-matrix. Coupled linear motion in two dimensions
does not include such restrictions and is in this sense a richer theory.

Nevertheless let us emphasise that the theory of symplectic motion in two dimensions does 
not allow scalar source terms, i.e. non-vanishing coefficients of $\y_{15}$ in 
the force matrix. If the MWEQs are derived from two-dimensional symplectic motion, 
then the Lorentz gauge is included. To make this clearer, we look at the first part of 
EQ.~\ref{eq_maxwell}, which includes a scalar expression $\d_t\phi+\vec\nabla\vec A$:
\begary{rcl}
{\bf F}&=&-{\bf D}\,{\bf\Phi}=-(\d_t\,\y_0-\vec\y\vec\nabla)\,(\phi\,\y_0+\vec\y\,\vec A)\\
                &=&\d_t\phi+\vec\nabla\vec A-(\d_t\vec A+\vec\nabla\phi)\,\y_0\,\vec\y\\
                &+&(\vec\nabla\times\vec A)\,\y_{14}\y_0\vec\y\\
                &=&(\d_t\phi+\vec\nabla\vec A)\,{\bf 1}+\vec E\,\y_0\,\vec\y+\vec B\,\y_{14}\y_0\vec\y\,,
\endary
where $\vec\y=(\y_1,\y_2,\y_3)^T$.
Since the force matrix may be composed exclusively of symplices, the scalar term has to vanish, i.e.:
\begeq
 \d_t\phi+\vec\nabla\vec A=0\,,
\endeq
which is the so-called ``Lorentz gauge''. In Sec.~\ref{sec_DualityRot} it will be shown, 
that the so-called {\it duality rotation} also has properties which are incompatible to 
the presented theory of two-dimensional symplectic motion.

In the next section we give a survey of symplectic transformations using the EMEQ
according to Eq.~\ref{eq_edeq}.
This concept allows to obtain a very transparent and systematic treatment 
of two-dimensional coupled linear optics. 
Of course there is some danger of puzzling the {\it formally introduced} 
electromagnetic components of the EMEQ with the fields of the accelerator 
due to the simi\-larity of notation.

\subsection{A Survey of Symplectic Transformations}
\label{sec_symplectic_trafo}

Symplectic transformations have been introduced by EQs.~\ref{eq_linear_eqom} 
and~\ref{eq_transfer_matrix}. It was shown that the exponentials
of symplices are symplectic and that there are ten symplices (in two dimensions)
corresponding to ten free parameters in symplectic matrices. 
This indicates a one-to-one relation between the symplices and the corresponding 
symplectic matrix. In the following we will classify the transformations that are 
based on single $\y$-matrices. Six of these transformations are known
as {\it rotations} and {\it Lorentz boosts} in 3-dimensional space,
based on the six bi-vectors $\y_4\dots\y_9$, i.e. the ``electric and magnetic 
field components''.
We call the four remaining transformations {\it phase rotation} (induced by
$\y_0$, i.e. by ``energy'') and {\it phase boosts} induced by the spatial 
basis vectors $\y_1\dots\y_3$ (``momentum''), respectively. 
All transformations are controlled by a continuous parameter $a$ 
such that the inverse transformation is given by the negative
of this parameter and a vanishing parameter $a=0$ yields the unit matrix:
\begary{rcl}
{\bf R}(0)&=&{\bf 1}\\
{\bf R}(a)^{-1}&=&{\bf R}(-a)\,.
\endary
The parameter $a$ has a specific physical meaning in the EMEQ. In case of 
rotations it is the angle, in case of Lorentz boosts it is the inverse hyperbolic 
tangent of $\beta=v/c$. 

Before discussing the action of the individual $\y$-symplices, we look again at the 
Hamiltonian in order to show exactly why the transformation matrix should
be symplectic:
\begeq
H(p,q)=\psi^T\,{\bf A}\,\psi=-\psi^T\,\y_0\,{\bf F}\,\psi\,,
\endeq
and insert the equation for a symplectic transformation of the forces ${\bf F'}={\bf R}\,{\bf F}\,{\bf R}^{-1}$, then
we find (using $\bar\psi\equiv\psi^T\,\y_0$)
\begary{rcl}
H(p,q)&=&-\bar\psi\,{\bf R}^{-1}\,{\bf R}\,{\bf F}\,{\bf R}^{-1}\,{\bf R}\,\psi\\
      &=&-(\bar\psi\,{\bf R}^{-1})\,({\bf R}\,{\bf F}\,{\bf R}^{-1})\,({\bf R}\,\psi)\\
      &=&(\bar\psi\,{\bf R}^{-1})\,\y_0\,\y_0\,({\bf R}\,{\bf F}\,{\bf R}^{-1})\,({\bf R}\,\psi)\\
      &=&-(\bar\psi\,{\bf R}^{-1}\,\y_0)\,{\bf A}'\,\psi'\\
      &=&\psi'^T\,{\bf A}'\,\psi'\,,
\endary
that the Hamiltonian is invariant under Lorentz transformations (LT) and under the assumption that
the spinors transform according to
\begary{rcl}
{\bf F}'&=&{\bf R}\,{\bf F}\,{\bf R}^{-1}\\
\psi'&=&{\bf R}\,\psi\\
\psi'^T&=&-\psi^T\,\y_0\,{\bf R}^{-1}\,\y_0=\psi^T\,{\bf R}^T\\
\Rightarrow&&-\y_0\,{\bf R}^{-1}\,\y_0={\bf R}^T\\
\Rightarrow&&\y_0={\bf R}^T\,\y_0\,{\bf R}\,.
\label{eq_symtransform}
\endary
or -- in other words: If a transformation matrix ${\bf R}$ is symplectic,
then the form (and the value) of the Hamiltonian and of the EQOM is conserved. 
Note that if the force matrix is transformed according to Eq.~\ref{eq_symtransform}, then
we find for the transfer matrix:
\begary{rcl}
{\bf M}'&=&\exp{({\bf F}'\,s)}=\sum\limits_{k=0}^\infty\,{({\bf F}'\,s)^k\over k!}\\
        &=&\sum\limits_{k=0}^\infty\,{({\bf R}\,{\bf F}\,{\bf R}^{-1}\,s)^k\over k!}\\
        &=&{\bf R}\,\left(\sum\limits_{k=0}^\infty\,{({\bf F}\,s)^k\over k!}\right){\bf R}^{-1}\\
        &=&{\bf R}\,{\bf M}\,{\bf R}^{-1}\,.
\label{eq_Msymtrans}
\endary

Symplectic transformation matrices have the form
\begary{lcl}
{\bf R}&=&\exp{({\bf F}\,\eps)}\\
{\bf R}^{-1}&=&\exp{(-{\bf F}\,\eps)}\,,
\endary
where the matrix ${\bf F}$ is a symplex.
The effect of a basic symplex $\y_b$ is given by:
\begary{rcl}
{\bf R}_b&=&\exp{(\y_b\,\eps)}\\
{\bf R}_b&=&\sum\limits_{k=0}^\infty\,{(\y_b\,\eps)^{2\,k}\over (2\,k)!}+\sum\limits_{k=0}^\infty\,{(\y_b\,\eps)^{2\,k+1}\over (2\,k+1)!}\\
         &=&\sum\limits_{k=0}^\infty\,(\y_b^2)^k\,{\eps^{2\,k}\over (2\,k)!}+\y_b\,\sum\limits_{k=0}^\infty\,(\y_b^2)^k\,{\eps^{2\,k+1}\over (2\,k+1)!}\\
         &=&\left\{
\begin{array}{lcl}
{\bf 1}\,\cos{(\eps)}+\y_b\,\sin{(\eps)}&\mathrm{for}&\y_b^2=-1\\
{\bf 1}\,\cosh{(\eps)}+\y_b\,\sinh{(\eps)}&\mathrm{for}&\y_b^2=1\\
\end{array}\right.\\
{\bf R}_b^{-1}&=&\exp{(-\y_b\,\eps)}\\
       &=&\left\{
\begin{array}{lcl}
{\bf 1}\,\cos{(\eps)}-\y_b\,\sin{(\eps)}&\mathrm{for}&\y_b^2=-1\\
{\bf 1}\,\cosh{(\eps)}-\y_b\,\sinh{(\eps)}&\mathrm{for}&\y_b^2=1\\
\end{array}\right.\\
\label{eq_sine_cosine}
\endary
Transformations with $\y_b^2=-1$ are orthogonal transformations, i.e. {\it rotations}, while those with $\y_b^2=1$ are {\it boosts}.
According to Tab.~\ref{tab_gamma}, $\y_b$ with $b\in [0,7,8,9]$ produce rotations and $\y_b$ with $b\in [1\dots 6]$ boosts.
Hence the transformed matrices are
\begary{rcl}
\y_a'&=&{\bf R}^{-1}\,\y_a\,{\bf R}\\
       &=&{\bf R}^{-1}\,\y_a\,{\bf R}\\
       &=&(c-s\,\y_b)\,\y_a\,(c+s\,\y_b)\\
       &=&c^2\,\y_a-\y_b\,\y_a\,\y_b\,s^2+c\,s\,(\y_a\,\y_b-\y_b\,\y_a)\,,
\endary
where $c$ and $s$ are the (hyperbolic) sine- and cosine-functions according to Eq.~\ref{eq_sine_cosine}.
The last term on the right vanishes if $\y_b$ and $\y_a$ commute. In this case the matrix $\y_a$ remains unchanged:
\begary{rcl}
\y_a'&=&\y_a\,(c^2-\y_b^2\,s^2)\\
       &=&\y_a\,\left\{\begin{array}{lp{5mm}l}\cos^2{(\eps)}+\sin^2{(\eps)}=1&for&\y_b^2=-1\\
       \cosh^2{(\eps)}-\sinh^2{(\eps)}=1&for&\y_b^2=1\\
\end{array}\right.
\endary
If the RDMs $\y_b$ and $\y_a$ anticommute, one finds:
\begary{rcl}
\y_a'&=&\y_a\,(c^2+\y_b^2\,s^2)+2\,c\,s\,\y_a\,\y_b\\
       &=&\left\{\begin{array}{lp{5mm}l}
\y_a\,\cos{(2\,\eps)}+\y_a\,\y_b\,\sin{(2\,\eps)}&for&\y_b^2=-1\\
\y_a\,\cosh{(2\,\eps)}+\y_a\,\y_b\,\sinh{(2\,\eps)}&for&\y_b^2=1\\
\end{array}\right.
\endary
The elementary symplectic transformations (in 2 dimensions) are summarized in Tab.~\ref{tab_symtrans}.
The ``frequency doubling'' is an indication that the transformed elements are according to their transformation
properties second moments. This is usually taken into account by using a transformation with the ``half-angle''.

Note that even though we describe here the transformation of the matrices instead of the RDM-coefficients,
this is equivalent to the transformation of the RDM-coefficients assuming that the RDMs have the same
form in all coordinate systems. This is only a matter of notation. 

\begin{table}
{\tiny\begin{tabular}{|c|c|c|c|c|c|c|c|c|c|c|c|}\hline
                     & $\y_0$& $\y_1$&  $\y_2$&  $\y_3$&  $\y_4$ &  $\y_5$&  $\y_6$&  $\y_7$&  $\y_8$&  $\y_9$ \\\hline\hline
 $c=$                & $\cos{\eps}$& $\cosh{\eps}$& $\cosh{\eps}$& $\cosh{\eps}$& $\cosh{\eps}$& $\cosh{\eps}$& $\cosh{\eps}$& $\cos{\eps}$& $\cos{\eps}$&$\cos{\eps}$\\\hline
 $s=$                & $\sin{\eps}$& $\sinh{\eps}$& $\sinh{\eps}$& $\sinh{\eps}$& $\sinh{\eps}$& $\sinh{\eps}$& $\sinh{\eps}$& $\sin{\eps}$& $\sin{\eps}$&$\sin{\eps}$\\\hline
 $\y_0'=c\y_0..$&        &$+s\y_4$& $+s\y_5$& $+s\y_6$& $-s\y_1$ & $-s\y_2$& $-s\y_3$&        &        &         \\\hline
 $\y_1'=c\y_1..$& $-s\y_4$&       & $+s\y_9$& $-s\y_8$& $-s\y_0$ &        &        &        & $-s\y_3$& $+s\y_2$ \\\hline
 $\y_2'=c\y_2..$& $-s\y_5$&$-s\y_9$&        & $+s\y_7$&         & $-s\y_0$&        & $+s\y_3$&        & $-s\y_1$ \\\hline
 $\y_3'=c\y_3..$& $-s\y_6$&$+s\y_8$& $-s\y_7$&        &         &        & $-s\y_0$& $-s\y_2$& $+s\y_1$&         \\\hline
 $\y_4'=c\y_4..$& $+s\y_1$&$+s\y_0$&        &        &         & $+s\y_9$& $-s\y_8$&        & $-s\y_6$&  $\y_5$ \\\hline
 $\y_5'=c\y_5..$& $+s\y_2$&       & $+s\y_0$&        &  $-s\y_9$&        & $+s\y_7$& $+s\y_6$&        & $-s\y_4$ \\\hline
 $\y_6'=c\y_6..$& $+s\y_3$&       &        & $+s\y_0$&  $+s\y_8$& $-s\y_7$&        & $-s\y_5$& $+s\y_4$&         \\\hline
 $\y_7'=c\y_7..$&        &       & $-s\y_3$& $+s\y_2$&         & $-s\y_6$& $+s\y_5$&        & $-s\y_9$& $+s\y_8$ \\\hline
 $\y_8'=c\y_8..$&        &$+s\y_3$&        & $-s\y_1$&  $+s\y_6$&        & $-s\y_4$& $+s\y_9$&        & $-s\y_7$ \\\hline
 $\y_9'=c\y_9..$&        &$-s\y_2$& $+s\y_1$&        &  $-s\y_5$& $+s\y_4$&        & $-s\y_8$& $+s\y_7$&         \\\hline
\end{tabular}}
\caption{Table of symplectic transformations in 2 dimensions. $a$ indicates the rows and $b$ the column:
$\y_a'=\exp{(-\y_b\,{\eps/2})}\,\y_a\,\exp{(-\y_b\,{\eps/2})}$. If $\y_a$ and $\y_b$ anticommute, then the 
result is $\y_a'=c\,\y_a+s\,\y_a\,\y_b$ where $c$ and $s$ are the sine- and cosine-function of $\eps$,
if $\y_b^2=-1$, and hyperbolic
sine- and cosine-function if $\y_b^2=1$, respectively. If $\y_a$ and $\y_b$ commute, then $\y_a'=\y_a$.
The table follows the commutator-table (Tab.~\ref{tab_commtab}). 
\label{tab_symtrans}}
\end{table}

\subsection{Rotations and Lorentz Boosts }
\label{sec_LorentzBoost}

Lorentz boosts are all transformations induced by $\y_b$ with $b\in [4\dots 6]$, i.e. by the electric field
terms, while rotations are induced by $\y_b$ with $b\in [7\dots 9]$, i.e. by magnetic field terms.
Note that both transformations are induced by bi-vectors and not by the basic matrices.
Both - rotations and Lorentz boosts - are well-known~\cite{Good} and the matrix coefficients transform exactly
as the physical quantities associated by the EMEQ. Hence we do not need to repeat the formulas here
as all relevant properties are summarized in Tab.~\ref{tab_symtrans}. 

Nevertheless it is instructive to look at the rotation matrices with respect to the transformation
of the spinor $\psi$. The matrix of the spatial rotation about the $z$-axis, induced by $\y_9$,
is explicitely given by:
\begeq
\exp{(\y_9\,{\eps\over 2})}=\bmtx{cccc}
\cos{\eps\over 2}&0&-\sin{\eps\over 2}&0\\
0&\cos{\eps\over 2}&0&-\sin{\eps\over 2}\\
\sin{\eps\over 2}&0& \cos{\eps\over 2}&0\\
0&\sin{\eps\over 2}& 0&\cos{\eps\over 2}\\
\emtx\,.
\endeq
Obviously $\y_9$ induces a rotation in the plane of $q_1$ and $q_2$. The canonical momenta are rotated 
accordingly. Note that the rotation angle in the associated space is $\eps$, the corresponding angle 
in the plane of the canonical variables is half of that: $\eps/2$.

The rotation about the $x$-axis ($\y_7$) mixes the canonical coordinates and momenta $q_1$ and $p_2$
($p_1$ and $q_2$):
\begary{rcl}
\exp{(\y_7\,{\eps\over 2})}&=&\bmtx{cccc}
\cos{\eps\over 2}&0&0&\sin{\eps\over 2}\\
0&\cos{\eps\over 2}&-\sin{\eps\over 2}&0\\
0&\sin{\eps\over 2}&\cos{\eps\over 2}&0\\
-\sin{\eps\over 2}&0&0&\cos{\eps\over 2}\\
\emtx\,.
 \label{eq_rotx}
\endary
A {\it phase rotation} is a rotation in phase space, for instance a rotation in the plane
of $q_1$ and $p_2$ as described by Eq.~\ref{eq_rotx}. Another example is given by the 
rotation about the $y$-axis as induced by $\y_8$ - a combined phase rotation in opposite 
directions:
\begary{rcl}
\exp{(\y_8\,{\eps\over 2})}&=&\bmtx{cccc}
\cos{\eps\over 2}&\sin{\eps\over 2}&0&0\\
-\sin{\eps\over 2}&\cos{\eps\over 2}&0&0\\
0&0&\cos{\eps\over 2}&-\sin{\eps\over 2}\\
0&0&\sin{\eps\over 2}&\cos{\eps\over 2}\\
\emtx\,.
\endary

In combination with the phase rotation matrix $\y_0$, this gives:
\begary{rcl}
\exp{((\y_0+\y_8)\,{\eps\over 2})}&=&\bmtx{cccc}
\cos{\eps}&\sin{\eps}&0&0\\
-\sin{\eps}&\cos{\eps}&0&0\\
0&0&1&0\\
0&0&0&1\\
\emtx\\
\exp{((\y_0-\y_8)\,{\eps\over 2})}&=&\bmtx{cccc}
1&0&0&0\\
0&1&0&0\\
0&0&\cos{\eps}&\sin{\eps}\\
0&0&-\sin{\eps}&\cos{\eps}\\
\emtx\,,
\label{eq_phase_rot}
\endary
so that both degrees of freedom are here decoupled. Note that in this case, the rotation
angle $\eps$ is {\it not} divided by two. 

We call the remaining symplectic transformations {\it phase rotation} induced by $\y_0$ 
and {\it phase boosts} induced by $\y_1$, $\y_2$ and $\y_3$. This is not quite correct, 
since we saw that also $\y_8$ induces a phase rotation. Strictly speaking, there is only 
one ``real'' spatial rotation possible in two dimensions - and this is induced by $\y_9$. 
All other ``rotations'' are rotations in phase space and are therefore phase rotations. 
Nevertheless the rotations induced by $\y_7$ and $\y_8$ are spatial rotations in the 
context of the EMEQ.

\subsection{Phase Rotation}
\label{sec_PhaseRot}

The symplectic transformation driven by $\y_0$ is a ``rotation'' between spacelike components and 
electric field components, represented by $\y_0\,\vec\y$. $\y_0$ and the magnetic components
$\y_7$, $\y_8$, $\y_9$ are unchanged. Expressing this using EMEQ one may write:
\begary{rcl}
E'&=&E\\
\vec p'&=&\cos{\eps}\,\vec p-\sin{\eps}\,\vec E\\
\vec E'&=&\cos{\eps}\,\vec E+\sin{\eps}\,\vec p\\
\vec B'&=&\vec B\\
\endary
The rotation matrix is given by
\begary{rcl}
\exp{(\y_0\,{\eps\over 2})}&=&\bmtx{cccc}
\cos{\eps\over 2}&\sin{\eps\over 2}&0&0\\
-\sin{\eps\over 2}&\cos{\eps\over 2}&0&0\\
0&0&\cos{\eps\over 2}&\sin{\eps\over 2}\\
0&0&-\sin{\eps\over 2}&\cos{\eps\over 2}\\
\emtx\,.
\endary
As listed in Tab.~\ref{tab_basis_sets}, the matrices $\y_0,\y_4,\y_5,\y_6$ form an alternative basis
to $\y_0,\y_1,\y_2,\y_3$. The phase rotation produced by $\y_0$ changes the ``mixing'' angle of the
standard basis and this first alternative basis. 

\subsection{Phase Boost}
\label{sec_PhaseBoost}

The transformations induced by $\y_1$, $\y_2$ and $\y_3$ are called ``phase boosts.'' 
These boosts are the exact analogue of the Lorentz boosts, but the transformation 
properties of $\vec E$ and $\vec p$ are exchanged, as can be seen from Tab.~\ref{tab_symtrans}. 
While Lorentz boosts are known to have the invariants $\vec E\,\vec B$ and $\vec E^2-\vec B^2$, 
a phase boost has the invariants $\vec p\,\vec B$ and $\vec p^2-\vec B^2$. 
Hence the phase boosts transform mass and electromagnetic energy into each other
as the Lorentz boosts are transforming mass and kinetic energy. We will come back to this 
in Sec.~\ref{sec_decoupling}.

\subsection{The Duality Rotation}
\label{sec_DualityRot}

The duality rotation can be expressed by the matrix $\y_{14}$, but it is not
a symplectic transformation and has some surprising and puzzling features. Since $\y_{14}$
commutes with all electric and magnetic field components (see Tab.~\ref{tab_commtab}),
the usual form of the transformation can not transform the fields, but only the particle
properties of energy and momentum. Using the nomenclature of Eq.~\ref{eq_edeq1}, 
one finds:
\begary{rcl}
{\bf R}&=&\exp{(\y_{14}\,\eps/2)}\\
{\bf R}^{-1}&=&\exp{(-\y_{14}\,\eps/2)}\\
{\bf F}_{ed}&=&{\bf R}\,{\bf F}_{ed}\,{\bf R}^{-1}={\bf R}^{-1}\,{\bf F}_{ed}\,{\bf R}\\
{\bf F}_{m}&=&{\bf R}\,{\bf F}_{m}\,{\bf R}\\
{\bf F'}_{m}&=&{\bf R}\,{\bf F}_{m}\,{\bf R}^{-1}=\cos{\eps}\,{\bf F}_m+\sin{\eps}\,\y_{14}\,{\bf F}_m\\
{\bf F'}_{ed}&=&{\bf R}\,{\bf F}_{ed}\,{\bf R}=\cos{\eps}\,{\bf F}_{ed}+\sin{\eps}\,\y_{14}\,{\bf F}_{ed}\\
\label{eq_duality}
\endary
The last line of Eqs.~\ref{eq_duality} is usually referred to as duality rotation,
explicitely given by:
\begary{rcl}
\vec E'&=&\cos{\eps}\,\vec E-\sin{\eps}\,\vec B\\
\vec B'&=&\cos{\eps}\,\vec B+\sin{\eps}\,\vec E\\
\endary
The duality rotation fails to fulfill all reasonable requirements in the context
of the EMEQ. First, it is not symplectic. Second, the transformation is not a
similarity transformation and finally, it has different forms for the mechanical
and the electrodynamical components.

If the MWEQs are deriveable from the EMEQ and symplectic flow, then there
is no magnetic charge or current density, but the MWEQs are forced 
to have the ``classical'' form with
\begary{rcl}
\vec\nabla\vec B&=&0\\
\vec\nabla\times\vec E+\d_t\,\vec B&=&0\\
\endary
To conclude - the duality rotation does not fit into in the concept of the EMEQ.

\subsection{Real Dirac Matrices and Elements in Coupled Linear Optics}
\label{sec_rdm_optics}

In the following we give some examples of force matrices as used in linear optics
of ion beam physics~\cite{Brown,Hinterberger}. 
We use the transversal coordinates $(x,x',y,y')$ as the pairs of 
conjugate variables $(q_1,p_1,q_2,p_2)$.
In ion beam physics there is the convention to use the path length $s$ along the
reference orbit as the independent variable.
The force matrix of a drift (force free motion) is given by
{\small\begeq
\dot\psi={d\over ds}\,\bmtx{c}x\\x'\\y\\y'\emtx=\bmtx{cccc}0&1&0&0\\0&0&0&0\\0&0&0&1\\0&0&0&0\emtx\,\bmtx{c}x\\x'\\y\\y'\emtx={\y_0+\y_6\over 2}\,\psi\,.
\label{eq_drift_mtx}
\endeq}

The transversal terms in a solenoid field:
{\small\begary{rcl}
{d\over ds}\,\bmtx{c}x\\x'/K\\y\\y'/K\emtx=K\,\bmtx{cccc}
0&1&1&0\\
-1&0&0&1\\
-1&0&0&1\\
0&-1&-1&0\emtx\,\bmtx{c}x\\x'/K\\y\\y'/K\emtx\,,
\endary}
where $K={B_S\over 2\,(B\,\rho)}$ with the solenoid field $B_S$,
charge $q$ and bending radius $\rho$. The bending radius depends
on the momentum $p$ and is given by
\begeq
\rho={p\over q\,B}\,.
\endeq
This can be written using RDMs as:
\begeq
\dot\psi=K\,(\y_0-\y_9)\,\psi\,,
\endeq
A radially focussing quadrupole is described by:
{\small\begary{rcl}
{d\over ds}\,\bmtx{c}x\\x'/K\\y\\y'/K\emtx=K\,\bmtx{cccc}
0&1&0&0\\
-1&0&0&0\\
0&0&0&1\\
0&0&1&0\emtx\,\bmtx{c}x\\x'/K\\y\\y'/K\emtx\,,
\endary}
where $K^2={\vert g\vert\over (B\rho)}$ with $g={\d B_y\over\d x}={\d B_x\over\d y}$ so that
-- written again with RDMs -- one finds:
\begeq
\dot\psi=K\,{\y_0+\y_1+\y_6+\y_8\over 2}\,\psi\,.
\endeq
The axially focusing quad is given by
\begeq
\dot\psi=K\,{\y_0-\y_1+\y_6-\y_8\over 2}\,\psi\,.
\endeq
A horizontal bending magnet: 
{\small\begary{rcl}
{d\over ds}\,\bmtx{c}x\\\rho x'\\y\\\rho y'\emtx&=&{1\over\rho}\bmtx{cccc}
0&1&0&0\\
-1&0&0&0\\
0&0&0&1\\
0&0&0&0\emtx\,\bmtx{c}x\\\rho x'\\y\\\rho y'\emtx\,,
\endary}
expressed by RDMs:
\begeq
\dot\psi={1\over\rho}\,{3\,\y_0+\y_1+\y_6+\y_8\over 4}\,\psi\,.
\endeq
Obviously the scale of the momentum coordinates has to be adjusted 
to all individual element types and changes even with their excitation (and the
beam energy), while the scale of the position coordinates is fix.
The scaling transformations that are required have the following form 
{\small\begary{rcl}
\bmtx{c}a\,x\\b\,x'\\a\,y\\b\,y'\emtx&=&\bmtx{cccc}
a&0&0&0\\
0&b&0&0\\
0&0&a&0\\
0&0&0&b\emtx\,\bmtx{c}x\\ x'\\y\\ y'\emtx\,,
\endary}
which allows to scale coordinates and momenta separately.
This matrix -- let us call it ${\bf R}$ -- can be written as 
\begeq
{\bf R}={a+b\over 2}\,{\bf 1}-{a-b\over 2}\,\y_3\,,
\endeq
and it is symplectic, if
{\small\begary{rcl}
{\bf R}^T\,\y_0\,{\bf R}&=&\left({a+b\over 2}\,{\bf 1}-{a-b\over 2}\,\y_3\right)\,\y_0\,\left({a+b\over 2}\,{\bf 1}-{a-b\over 2}\,\y_3\right)\\
                        &=&\left({(a+b)^2\over 4}-{(a-b)^2\over 4}\right)\,\y_0\\
                        &=&a\,b\,\y_0\\
\Rightarrow             && a\,b=1\,,
\endary}
or -- with the equivalent definition
{\small\begary{rcl}
{\bf R}&=&\exp{(\y_3\,\eps)}=\bmtx{cccc}
e^{-\eps}&0&0&0\\
0&e^\eps&0&0\\
0&0&e^{-\eps}&0\\
0&0&0&e^\eps\emtx\\
\endary}
Hence the phase boost in the direction $\y_3$ can be understood as a transformation 
that changes the relative scale of the classical coordinates and momenta. The other scaling 
transformations are induced by the other diagonal RDMs. In the chosen basis these
are $\y_4$, $\y_{12}$ and $\y_{15}={\bf 1}$. Besides $\y_3$, only $\y_4$ is a symplex and 
is a legitimate candidate for a symplectic scaling transformation:
\begary{rcl}
{\bf R}&=&\exp{(\y_4\,\eps)}\\
       &=&\mathrm{Diag}(e^{-\eps},e^{\eps},e^{\eps},e^{-\eps})\\
\label{eq_scaling_trafo}
\endary
while $\y_{12}$ and $\y_{15}$ do not generate symplectic transformations:
\begary{rcl}
\exp{(\y_{12}\,\eps)} &=&\mathrm{Diag}(e^{-\eps},e^{-\eps},e^{\eps},e^{\eps})\\
\exp{(\y_{15}\,\eps)} &=&\mathrm{Diag}(e^{\eps},e^{\eps},e^{\eps},e^{\eps})\\
\endary

We note that there are symplectic transformations to independently change the scales between 
$q_1$ and $p_1$ and between $q_2$ and $p_2$, respectively. But there is no symplectic 
transformation to change the scales between the directions $(q_1,p_1)$ and $(q_2,p_2)$.
Nevertheless, the rescaling that is required to compare force matrices
of quadrupole, solenoids, dipols, etc. to RDMs can be done by symplectic
transformations, namely by ${\bf R}_3=\exp{(\y_3\,\eps)}$.

\section{Decoupling 2-Dim. Harmonic Oscillators}
\label{sec_decoupling}

Consider a system of two coupled degrees of freedom as it is common in many parts of
physics. In the general case, the force matrix will be time dependent and so will be
the symplectic transformation matrix ${\bf R}$:
\begary{rcl}
\tilde\psi&=&{\bf R}\,\psi\\
\dot{\tilde\psi}&=&{\bf\dot R}\,\psi+{\bf R}\,\dot\psi\\
              &=&\left({\bf\dot R}+{\bf R}\,{\bf F}\right)\,\psi\\
              &=&\left({\bf\dot R}\,{\bf R}^{-1}+{\bf R}\,{\bf F}\,{\bf R}^{-1}\right)\,\tilde\psi={\bf\tilde F}\,\tilde\psi\\
\label{eq_symtrans_t}
\endary
Decoupling now means that the transformed force matrix 
\begeq
{\bf \tilde F}={\bf\dot R}\,{\bf R}^{-1}+{\bf R}\,{\bf F}\,{\bf R}^{-1}
\label{eq_forcetrans}
\endeq
is either a SYSY or block-diagonal. It may still be time-dependent.
Eq.~\ref{eq_forcetrans} can also be written as~\cite{Leach} 
\begeq
{\bf\dot R}={\bf \tilde F}\,{\bf R}-{\bf R}\,{\bf F}\,.
\endeq
Given that the transformation matrix has the form
\begeq
{\bf R}=\exp{(-{\bf G}\,\tau)}\,,
\endeq
where ${\bf G}$ is constant, then one obtains
\begary{rcl}
{\bf\dot R}&=&-{\bf G}\,{\bf R}\\
{\bf\tilde F}&=&-{\bf G}+{\bf R}\,{\bf F}\,{\bf R}^{-1}\\
\label{eq_ttrans}
\endary
If ${\bf\tilde F}=\mathrm{const}$, then it follows from Eq.~\ref{eq_ttrans}
and from the fact that ${\bf R}$ and ${\bf G}$ commute, that
\begary{rcl}
\dot{\bf \tilde F}&=&\dot{\bf R}\,{\bf F}\,{\bf R}^{-1}+{\bf R}\,\dot{\bf F}\,{\bf R}^{-1}+{\bf R}\,{\bf F}\,\dot{\bf R}^{-1}=0\\
0                 &=&-{\bf G}\,{\bf R}\,{\bf F}\,{\bf R}^{-1}+{\bf R}\,\dot{\bf F}\,{\bf R}^{-1}+{\bf R}\,{\bf F}\,{\bf G}\,{\bf R}^{-1}\\
\dot{\bf F}       &=&{\bf G}\,{\bf F}-{\bf F}\,{\bf G}\,.
\label{eq_dotforce}
\endary
Obviously, the form of the Lorentz force equation (Eq.~\ref{eq_lorentzforce}) is very
common in coupled linear optics and can be obtained with a few simple assumptions 
for symplectic transformation matrices as in Eq.~\ref{eq_forcetrans} or for symplices
as in Eq.~\ref{eq_dotforce} or for second moments (i.e. the ${\bf S}$-matrix, respectively)
as in Eq.~\ref{eq_dotsigma}.

In the following we present a straightforward recipe for decoupling {\it constant} 
force matrices. It will be shown that this method can also be applied to transfer matrices -
hence can be used to compute the properties of matched beams even in cases where the
force matrix itsself is not constant.

At the end of this section we scetch some problems related to the general case of 
time-dependent forces. But a comprehensive treatment of the time-dependent case
is beyond the scope of this article.

\subsection{Eigenvalues of the force matrix}
\label{sec_Feigen}

The eigenvalues of the force matrix as expressed by Eq.~\ref{eq_edeq1} are given by
\begary{rcl}
\lambda&=&\mathrm{Diag}(i\,\omega_1,-i\,\omega_1,i\,\omega_2,-i\,\omega_2)\\
K_1&=&{\cal E}^2+\vec B^2-\vec E^2-\vec p^2\\
K_2&=&-2\,{\cal E}\,\vec p\,(\vec E\times\vec B)+{\cal E}^2\,\vec B^2+\vec E^2\,\vec p^2\\
   &-&(\vec E\vec p)^2-(\vec E\vec B)^2-(\vec p\vec B)^2\\
\omega_1 &=&\sqrt{K_1+2\,\sqrt{K_2}}\\
\omega_2 &=&\sqrt{K_1-2\,\sqrt{K_2}}\\
\mathrm{Det}({\bf F})&=&K_1^2-4\,K_2\\
\label{eq_eigenfreq}
\endary
The system is stable, if the eigenvalues are purely imaginary, i.e. if the
eigen{\it frequencies} are real. This is the case, if the following conditions 
are fullfilled:
\begary{rcl}
K_2&\ge&0\\
K_1&\ge&2\,\sqrt{K_2}\\
\endary
The force matrix can only be a SYSY, if the eigenfrequencies are equal, i.e. if
$K_2=0$. This is for instance the case, if $\vec E=\vec B=0$. If the vectors
$\vec p$, $\vec E$ and $\vec B$ are orthogonal to each other such that
$\vec p=\alpha^2\,\vec E\times\vec B$ with some constant $\alpha$, then it follows
\begary{rcl}
K_2&=&-2\,{\cal E}\,\vec p\,(\vec E\times\vec B)+{\cal E}^2\,\vec B^2+\vec E^2\,\vec p^2\\
   &=&-2\,{\cal E}\,p\,E\,B+{\cal E}^2\,B^2+E^2\,p^2\\
   &=&({\cal E}\,B-E\,p)^2\,,
\label{eq_K2ortho}
\endary
where $E=\vert\vec E\vert$, $B=\vert\vec B\vert$ and $p=\vert\vec p\vert$.
In this case the force matrix is a SYSY, if ${\cal E}\,B=E\,p$.

We will show in the following that the standard form of the block-diagonal 
force matrix as it can be obtained by symplectic transformations, is given by
\begary{rcl}
{\bf F}_d&=&\bmtx{cccc}
0                    &\alpha  &  0  & 0\\
-\beta               & 0                &  0  & 0\\
0                    & 0                &  0  &\gamma\\
0                    & 0                &  -\delta  &    0   \\
\emtx\\
\omega_1&=&\pm\,\sqrt{\alpha\,\beta}\\
\omega_2&=&\pm\,\sqrt{\gamma\,\delta}\,.
\label{eq_F_std}
\endary
The Hamilton matrix ${\bf A}=-\y_0\,{\bf F}_d$ corresponding to this force matrix 
is diagonal:
\begary{rcl}
{\bf A}&=&-\y_0\,{\bf F}_d\\
&=&\mathrm{Diag}(\beta,\alpha,\delta,\gamma)\\
\label{eq_basic_hamilton}
\endary
The RDM-composition of ${\bf F}_d$ is given by
\begary{rcl}
{\bf F}_d&=&{\cal E}\,\y_0+B_y\,\y_8+p_x\,\y_1+E_z\,\y_6\\
{\cal E}&=&{\alpha+\beta+\gamma+\delta\over 4}\\
B_y&=&{\alpha+\beta-\gamma-\delta\over 4}\\
p_x&=&{-\alpha+\beta+\gamma-\delta\over 4}\\
E_z&=&{\alpha-\beta+\gamma-\delta\over 4}\\
\endary
The vectors $\vec p$, $\vec E$ and $\vec B$ are orthogonal to each other
so that Eq.~\ref{eq_K2ortho} holds, but with a reversed sign for the triple product.
Obviously one can exchange ${\cal E}$ with $B_y$ and $p_x$ with $E_z$ without changing the
eigenfrequencies. This proves that stable systems are possible with $B_y>{\cal E}$ and/or
with $E_z>p_x$, i.e. with $\vert\vec B\vert>{\cal E}$ and /or $\vert\vec E\vert>\vert\vec P\vert$.

We define a stable system to be {\it regular} or {\it massive}, if all parameters 
$\alpha$, $\beta$, $\gamma$ and $\delta$ have the same sign (are positive), i.e. if the decoupled Hamiltonian 
function in the new coordinates $\tilde H=\tilde\psi^T\,{\bf\tilde A}\,\tilde\psi$ is positive definite:
\begeq
\tilde H=\alpha\,\tilde p_1^2+\beta\,\tilde q_1^2+\gamma\,\tilde p_2^2+\delta\,\tilde q_2^2\,.
\endeq
In a regular system the coefficients of the force matrix according to Eq.~\ref{eq_F_std} yield
\begary{rcl}
{\cal E}^2&>& B_y^2 \\
{\cal E}^2&>& p_x^2 \\
{\cal E}^2&>& E_z^2 \,.
\endary
If one pair, either $(\alpha,\beta)$ or $(\gamma,\delta)$ is negative, then the Hamiltonian function
contains a negative kinetic energy term and is indefinite.
We then call the system {\it irregular} or {\it magnetic} and it is
easily shown that in this case
\begary{rcl}
B_y^2 &>&{\cal E}^2\\ 
B_y^2 &>& p_x^2 \\
B_y^2 &>& E_z^2 \,.
\endary

\subsection{Decoupling a Constant Focussing Channel in two Dimensions}
\label{sec_decoupling_const}

The RDMs and the presented theory of symplectic transformations together with 
the EMEQ facilitates the task of decoupling - since we have already the 
{\it geometrical and physical understanding} that provides the optimal strategy 
to solve the problem. In Sec.~\ref{sec_coupling_def} we discussed the meaning
of coupling and it turned out that constant symplectic force matrices (SYSYs) 
do not require decoupling. Instead the next derivative yields a decoupled
second-order equation. In the following we would like to discuss decoupling
of EQ.~\ref{eq_linear_eqom} in cases where the force matrix is not a SYSY.
Then decoupling is the task to find a transformation matrix ${\bf R}$ such that
\begary{rcl}
{\bf R}\,{\bf F}\,{\bf R}^{-1}&=&\bmtx{cc} {\bf A}& 0\\0&{\bf B}\emtx\,.
\endary
Since products of symplectic transformations are again symplectic, we
may find the solution in several steps.

One distinguishes even and odd matrices, where odd matrices couple the two degrees 
of freedom while even ones do not~\cite{Foldy}.
In the chosen basis eight matrices, namely $\y_0$, $\y_1$, $\y_3$, $\y_4$, $\y_6$, $\y_8$, 
$\y_{12}$, $\y_{15}$ are even and the remaining eight matrices are odd.
Since the transformation is required to be symplectic, the anti-symplices 
$\y_{10}\dots\y_{15}$ can be excluded.
There are six even  ($\y_0$, $\y_1$, $\y_3$, $\y_4$, $\y_6$ and $\y_8$) and
four odd matrices ($\y_2$, $\y_5$, $\y_7$ and $\y_9$) remaining.
This can directly be seen from the force matrix, which is given by
\begary{rcl}
{\bf F}&=&{\cal E}\,\y_0+\vec p\,\vec\y+\vec E\,\y_0\,\vec\y+\vec B\,\y_{14}\,\y_0\,\vec\y\\
       &=&{\tiny\bmtx{cc}{\bf A}_0 & \begin{array}{cc}
 -B_z+E_y       &  B_x+p_y \\
 -B_x+p_y       & -B_z-E_y \\
\end{array}\\
\begin{array}{cc}
 E_y+B_z       &        B_x+p_y \\
-B_x+p_y       & B_z-E_y        \\
\end{array}&{\bf B}_0
\emtx}
\endary
The odd components are $p_y$, $E_y$, $B_x$ and $B_z$ corresponding to the odd matrices
$\y_2$, $\y_5$, $\y_7$ and $\y_9$.

Products of even matrices result even matrices. Hence the transformation to block-diagonal 
form can be done by the exclusive use of odd force matrices. Note that the odd symplices 
all anti-commute with each other. Two of them ($\y_2$ and $\y_5$) square 
to $+{\bf 1}$, the other two to $-{\bf 1}$. Hence they are a basis of the 
Clifford-algebra $Cl(2,2)$.

Even matrices may be used for convenience, but exclusively odd matrices are {\it required}
to bring a regular force matrix into block-diagonal form. 
The transformations have the form of Eqs.~\ref{eq_symtransform} and~\ref{eq_sine_cosine},
where in the latter we replace the parameter $\tau$ with ${\eps_b\over 2}$:
\begary{rcl}
{\bf R}_b&=&\exp{(\y_b\,\eps_b/2)}\\
{\bf R}_b^{-1}&=&\exp{(-\y_b\,\eps_b/2)}\\
{\bf F}'&=&{\bf R}_b\,{\bf F}\,{\bf R}_b^{-1}\\
\endary 
After each transformation, the coefficients have to be updated according to Eqs.~\ref{eq_analyse} and~\ref{eq_scalarprod_def}.

Then the recipe for {\it regular} systems is as follows:
\begin{enumerate}
\item Use the rotation-matrices ${\bf R}_7$ (about $x$-axis with $\eps_7=\arctan{\left({p_z\over p_y}\right)}$) and
${\bf R}_9$ (about $z$-axis with $\eps_9=-\arctan{p_x\over p_y}$) to align the momentum $\vec p$ 
along the $y$-axis.
\item Use ${\bf R}_5$ with $\eps_5=-\mathrm{artanh}{p_y\over {\cal E}}$ (Lorentz boost) to transform into the rest frame.
\item Use again ${\bf R}_7$ and ${\bf R}_9$ (with  $\eps_7=\arctan{B_z\over B_y}$ and  $\eps_9=-\arctan{\left({B_x\over B_y}\right)}$) 
      to align the magnetic field along the $y$-axis. If this is done, we have $B_x=B_z=0$.  
\item Use ${\bf R}_8$ ($\eps_8=-\arctan{\left({E_x\over E_z}\right)}$) to rotate about $y$-axis and make $E_x$ vanish. Note that this step
      is not required to obtain block-diagonal form. But it is required for a complete diagonalization.
\item Use ${\bf R}_2$ ($\eps_2=\mathrm{artanh}{\left({E_y\over{\cal E}}\right)}$) to make the parallel electric field component $E_y$ vanish.
\end{enumerate}
All but the last step should be clear. In the rest frame we have $\vec p=0$, so that
rotation matrices can be applied without restrictions. The (odd) $\y_2$-term (i.e. $p_y$)
will remain zero. The alignment of $\vec B$ along the $y$-axis brings the odd terms
$B_x$ and $B_z$ to zero, the rotation about the $y$-axis yields $E_x\to 0$. 
The only remaining odd term is $\y_5$, i.e. $E_y$:
\begary{rcl}
{\bf F}^{(3)}&=&m\,\y_0+E_y\,\y_5+E_z\,\y_6+B_y\,\y_8\\
       &=&{\tiny\bmtx{cccc}
  0&B_y+E_z+m&E_y&0\\
-B_y+E_z-m&0&0&-E_y\\
E_y&0&0&-By+E_z+m\\
0&-E_y&B_y+E_z-m&0\\
\emtx}\,,
\endary
where we wrote $m$ instead of ${\cal E}$ as the $\y_0$ coefficient to make clear that the
transformation into the rest frame has already been done.
The last step is required to achieve the block-diagonal form of
Eq.~\ref{eq_F_std} where
\begary{rcl}
{\bf F}^{(4)}&=&\tilde m\,\y_0+E_z\,\y_6+B_y\,\y_8\\
\alpha&=&B_y+E_z+\tilde m\\
\beta &=&B_y-E_z+\tilde m\\
\gamma&=&-B_y+E_z+\tilde m\\
\delta&=&-B_y-E_z+\tilde m\\
\label{eq_basic_force}
\endary
where $\tilde m$ is given by
\begeq
\tilde m=\sqrt{m^2-E_y^2}\,.
\endeq
This shows that phase boosts are transformations which convert electromagnetic energy into mass and vice versa
in the same sense as Lorentz boosts convert mass into kinetic energy or vice versa.
Note that we could also exchange the order of the last two steps, i.e. of ${\bf R}_8$ and ${\bf R}_2$.
The result would be the same since $\y_8$ and $\y_2$ {\it commute}.

If the force matrix is {\it irregular} (but stable), then the energy term might be too small to perform 
the Lorentz boost into the rest frame as suggested by the recipe. In this case, the magnetic components
provide the focusing strength that stabilize the system (see Eq.~\ref{eq_eigenfreq}).

In this case we have to prepare the force matrix in the following way:
\begin{enumerate}
\item[1.] In the irregular case, we have to consider the possibility that $\vec p^2\ge{\cal E}^2$.
      In this case the first step is a phase rotation ${\bf R}_0$ with 
      $\eps_0={1\over 4}\,\arctan{\left({2\,\vec E\,\vec P\over \vec E^2-\vec P^2}\right)}$ 
      to minimize $\vec p^2$. 
\item[2.] Align $\vec B$ along the $y$-axis.
\item[3.] Rotate about $y$-axis to maximize $p_z$ (i.e. to make $p_x=0$).
\item[4.] Apply phase boost using $\y_1$ by $\eps_1={1\over 2}\,\mathrm{artanh}{({p_z\over B_y})}$.
\end{enumerate}
If the energy is still to low, perform the following steps:
\begin{enumerate}
\item[5.] Align (again) $\vec B$ along the $y$-axis.
\item[6.] Rotate about $y$-axis to maximize $E_z$ (i.e. to make $E_x=0$).
\item[7.] Apply Lorentz boost using $\y_4$ by $\eps_4={1\over 2}\,\mathrm{artanh}{({E_z\over B_y})}$.
\end{enumerate}
Now the energy ${\cal E}$ should fulfill ${\cal E}^2>\vec p^2$, so that the usual recipe for regular
matrices can be applied.

An inspection of Tab.~\ref{tab_basis_sets} may help to explain these transformations. Magnetic
force matrices are {\it irregular} with respect to the definition of the state vector, i.e.
the choice of $\y_0$ as explained above with Eq.~\ref{eq_stdvector} and Eq.~\ref{eq_alternatevector}.
If we inspect Tab.~\ref{tab_basis_sets}, then we can see that in basis system 9), where $\y_8$ 
represents the energy and $\y_3$ represents $p_x$, $\y_1$ is the driver of a Lorentz boost.
Interpreted in the usual system this means that it is possible to minimize $p_z$ using ${\bf R}_1$, 
if $B_y^2>p_z^2$. 
The second suggested transformation corresponds to the same basis and a Lorentz boost along $y$
(accordingly in basis system No. 9, this is driven by $\y_4$). The fact that we have to look at
a different basis system to find the appropriate symplectic transformation, legitimizes us to
claim that such systems are {\it irregular}. The examples given in Sec.~\ref{sec_rdm_optics} 
have $\y_0$-coefficients which are at least as strong as any other component. This fact seems 
to support the assumption that only regular systems exist in linear coupled optics. This assumption
is wrong and in the following section we give an example for an irregular system.

\subsection{Example for an Irregular System}
\label{sec_irregular}

In an accompanying paper we describe a simplified and idealized cyclotron model with space charge, which is
an example for an irregular system~\cite{cyc_paper}. 
The force matrix can easily be transformed into the form of Eq.~\ref{eq_basic_force}.
Using $\psi=(x,x',l,\delta)^T$ as the canonical coordinates, where $x$ and $l$ are the horizontal and
longitudinal position of the ion, $x'={dx\over ds}$ is the horizontal angle and $\delta={p-p_0\over p_0}$
is the relative deviation of the momentum $p$ from the average momentum $p_0$.
The (constant) force matrix ${\bf F}$ is given by
\begeq
{\bf F}=\bmtx{cccc}
0&1&0&0\\
-k_x+K_x&0&0&h\\
-h&0&0&{1\over\y^2}\\
0&0&K_z\,\y^2&0\\
\emtx\,,
\label{eq_eqom_sc}
\endeq
where $\y$ is the relativistic factor, $h=1/r$ is the inverse bending radius of the magnetic field
and $k_x=h^2\,(1+n)=h^2\,\y^2$ is the horizontal restoring force. $n={r\over B}\,{dB\over dr}$ 
is the field index. $K_x$ and $K_z$ are the horizontal and 
axial space charge force~\cite{cyc_paper}, respectively.
The eigenfrequencies of the force matrix are
\begary{rcl}
a&\equiv&{k_x-K_x-K_z\over 2}\\
b&\equiv&K_z\,(K_x+h^2\,\y^2-k_x)\\
\W&=&\sqrt{a+\sqrt{a^2-b}}\\
\w&=&\sqrt{a-\sqrt{a^2-b}}\,.
\label{eq_freq}
\endary
The RDM-coefficients are
\begary{rcl}
{\cal E}&=&{1\over 4}\,\left(1+k_x-K_x+{1\over\y^2}-\y^2\,K_z\right)\\
P_x&=&{1\over 4}\,\left(-1+k_x-K_x+{1\over\y^2}+\y^2\,K_z\right)\\
P_y&=&P_z=0\\
E_x&=&B_x=0\\
E_y&=&B_z=-{h\over 2}\\
E_z&=&{1\over 4}\,\left(1-k_x+K_x+{1\over\y^2}+\y^2\,K_z\right)\\
B_y&=&{1\over 4}\left(1+k_x-K_x-{1\over\y^2}+\y^2\,K_z\right)\\
\endary
The system is resonably simple as there are only two elements that are
not block-diagonal. This allows to guess a transformation matrix
instead of using the above recipe:
\begary{rcl}
{\bf R}&=&\exp{\left((r+1/r)\,{s\over 2}\,\y_2+(r-1/r)\,{s\over 2}\,\y_7\right)}\\
       &=&\cosh{(s)}\,{\bf 1}+\left((r+1/r)\,\y_2+(r-1/r)\,\y_7\right)\,{\sinh{(s)}\over 2}\\
\endary
Where the follwoing abbreviations are used:
\begary{rcl}
A&=&{h\over\W^2+K_z}\\ 
B&=&{h\over\w^2+K_z}\\
\cosh{(s)}&=&\sqrt{B\over B-A}\\
\sinh{(s)}&=&-\sqrt{A\over B-A}\\
r&=&{1\over\sqrt{K_z}\,\gamma}\\
\endary
The transformed force matrix is then
\begary{rcl}
{\bf\tilde F}&=&{\bf R}\,{\bf F}\,{\bf R}^{-1}\\
&=&\bmtx{cccc}
0&1&0&0\\
-\beta&0&0&0\\
0&0&0&-\Gamma\\
0&0&\y^2\,K_z&0\\
\emtx\\
\beta&=&{A\,K_z+B\,(k_x-K_x)-2\,h\over B-A}\\
\Gamma&=&{B\,K_z+A\,(k_x-K_x)-2\,h\over (B-A)\,\y^2\,K_z}\\
\endary
Since $\y^2\,K_z>0$ and $\Gamma>0$ is given, the transformed force matrix is
irregular and the transformed Hamiltonian $\tilde H$ has the form:
\begeq
{\tilde H}=\tilde x'^2+\beta\,\tilde x^2-\Gamma\,\tilde\delta^2-\y^2\,K_z\,l^2\,.
\endeq
It is also an example for a classical system, that can not be diagonalized by
the method of ``symplectic rotation'' of Teng and Edwards~\cite{cyc_paper}.

\subsection{Diagonalization}
\label{sec_diag}

Though the force matrix has been decoupled, there is motivation to continue the
process of diagonalization by a few more transformations, since then the product
of the transformation matrices equals the matrix of eigenvectors and the resulting
diagonal matrix contains the eigenvalues.
According to Eq.~\ref{eq_F_eigen},~\ref{eq_M_eigen},~\ref{eq_S_eigen}
and Eq.~\ref{eq_Sdiag}, the matched beam-ellipse can then be directly computed
for any combination of the emittances $\eps_1$ and $\eps_2$. 

The intention is to bring ${\bf F}$ into the form of EQ.~\ref{eq_F_eigen}.
Since we expect imaginary eigenvalues, the transformation matrices 
can not be real-symplectic.
The first ``direction'' that we use is $i\,\y_6$, which can formally be
identified with a Lorentz boost with an infinite imaginary relative ``velocity'',
so that the ``angle'' is given by
\begary{rcl}
\eps&=&\lim\limits_{\beta\to\infty}\,\mathrm{artanh}{(i\,\beta)}/2\\
   &=&\lim\limits_{\beta\to\infty}\,{i\over 2}\,\arctan{(\beta)}\\
   &=&i\,{\pi\over 4}\,.
\endary
The matrix is explicitely given by
\begary{rcl}
{\bf P}_6&=&\exp{(i\,\y_6\,{\pi\over 4})}\\
&=&{1\over\sqrt{2}}\,\bmtx{cccc}
1&-i&0&0\\
-i&1&0&0\\
0&0&1&-i\\
0&0&-i&1\\
\emtx\,.
\endary
This matrix has some remarkable properties: It is unitary, symmetric and symplectic.
The transformed force matrix is
\begary{rcl}
{\bf F}^{(6)}&=&{\bf P}_6\,{\bf F}^{(5)}\,{\bf P}_6^{-1}\\
        &=&{\tiny\bmtx{cccc}
i\,(B_y+\tilde m)&E_z&0&0\\
E_z&-i\,(B_y+\tilde m)&0&0\\
0&0&-i\,(B_y-\tilde m)&E_z\\
0&0&E_z&i\,(B_y-\tilde m)\\
\emtx}\\
\label{eq_diagonalize1}
\endary
And the final transformation includes two RDMs, $\y_0$ and $\y_8$. 
This combination allows for a relative phase rotation 
between both degrees of freedom as in EQ.~\ref{eq_phase_rot},
but now with imaginary angles:
\begary{rcl}
{\bf P}_8&=&\exp{(i\,\y_0\,{\eps_1-\eps_2\over 2}+i\,\y_8\,{\eps_1+\eps_2\over 2})}\\
&=&\bmtx{cccc}
c_1&i\,s_1&0&0\\
-i\,s_1&c_1&0&0\\
0&0&c_2&-i\,s_2\\
0&0&i\,s_2&c_2\\
\emtx\,,
\endary
where
\begary{rcl}
c_i&=&\cosh{\eps_i}\\
s_i&=&\sinh{\eps_i}\\
\eps_1&=&{1\over 2}\,\mathrm{Artanh}({E_z\over B_y+\tilde m})\\
\eps_2&=&{1\over 2}\,\mathrm{Artanh}({E_z\over B_y-\tilde m})\,,
\endary
so that
\begary{rcl}
{\bf F}^{(7)}&=&{\bf P}_8\,{\bf F}^{(6)}\,{\bf P}_8^{-1}=\lambda\\
        &=&{\tiny\bmtx{cccc}
i\,(B_y+\tilde m)\,a&0&0&0\\
0&-i\,(B_y+\tilde m)\,a&0&0\\
0&0&-i\,(B_y-\tilde m)\,b&0\\
0&0&0&i\,(B_y-\tilde m)\,b\\
\emtx}\\
a&=&\sqrt{1-{E_z^2\over(B_y+\tilde m)^2}}=\sqrt{4\,\alpha\beta\over(\alpha+\beta)^2}\\
b&=&\sqrt{1-{E_z^2\over(B_y-\tilde m)^2}}=\sqrt{4\,\delta\gamma\over(\delta+\gamma)^2}\\
\label{eq_diagonalize2}
\endary
In terms of RDMs the diagonalized force matrix is:
\begary{rcl}
{\bf F}^{(7)}
 &=&-{i\over 2}\,[a\,(B_y+\tilde m)-b\,(B_y-\tilde m)]\,\y_3\\
 &-&{i\over 2}\,[a\,(B_y+\tilde m)+b\,(B_y-\tilde m)]\,\y_4\\
\label{eq_diagonalize3}
\endary
Note that the last (double-) transformations is not necessary, if $E_z$ is
already zero. The eigenvectors are now given by the product of all
transformation matrices (in their order).
Since ${\bf R}_8$ is also symplectic, it turned out that the matrix of eigenvectors 
${\bf E}$ as a product of (complex) symplectic matrices is (complex) symplectic.

\subsection{Summary: Decoupling and Diagonalization}

The force matrix ${\bf F}$ of stable systems was shown to have the form
of Eq.~\ref{eq_F_eigen} where the diagonal matrix $\lambda$ has the form
of Eq.~\ref{eq_lambda}. It turned out that there is an intermediate stage
in the process which corresponds to a diagonalized Hamiltonian function.
The similarity transformation ${\bf R}$ to reach this intermediate stage is
symplectic and with Eq.~\ref{eq_Msymplectic2} and Eq.~\ref{eq_linear_eqom} 
it follows that
\begary{rcl}
{\bf A}_d&=&\mathrm{Diag}(\beta,\alpha,\delta,\gamma)\\
{\bf F}_d&=&{\bf R}\,{\bf F}\,{\bf R}^{-1}\\
\y_0\,{\bf A}_d&=&{\bf R}\,\y_0\,{\bf A}\,{\bf R}^{-1}\\
             &=&-{\bf R}\,\y_0\,{\bf A}\,\y_0\,{\bf R}^T\,\y_0\\
{\bf A}_d    &=&\y_0\,{\bf R}\,\y_0\,{\bf  A}\,\y_0\,{\bf R}^T\,\y_0\,.
\endary
This equation can be reversed by multiplication with ${\bf R}$ (${\bf R}^T$) from the right (left), respectively:
\begeq
{\bf  A}={\bf R}^T\,{\bf A}_d\,{\bf R}\,.
\endeq
Note that ${\bf R}$ is in general {\it not} orthogonal, ie. ${\bf R}^T\,{\bf R}\ne {\bf 1}$.
Hence a symplectic diagonalization can usually not be replaced by an orthogonal transformation.
Nevertheless the symplectic decoupling diagonalizes the Hamiltonian.

\subsection{The RDM-Coefficients of the Transfer Matrix and the Tunes}
\label{sec_M_compounds}

In the following we use the knowledge of the matrix of eigenvectors ${\bf E}$
to further investigate the general structure of the transfer matrix
using Eq.~\ref{eq_M_xeigen} and Eq.~\ref{eq_M_filter2}.
The ``amplitudes'' of ${\bf M}_s$ are limited by the sine-function, but the 
structure is identical to the structure of the force matrix, i.e. ${\bf M}_s$ is
composed of symplices. 
The structural difference between transfer and force matrix is given by ${\bf M}_c$.

Note that the decoupling algorithm of Teng and Edwards for twodimensional systems is applied 
to the transfer matrix instead of the force matrix. But from Eq.~\ref{eq_M_xeigen} and 
Eq.~\ref{eq_M_xcoeff} its it clear that the computation of the eigenvectors does not require 
to include ${\bf M}_c$. The decoupling can therefore be done with the force matrix ${\bf F}$,
- which is favorable in case of constant forces - or with ${\bf M}_s$. The latter enables
the treatment of periodic systems with varying force matrix ${\bf F}$. 
If the decoupling algorithm is applied to ${\bf M}_s$, then one obtains by comparison of
Eq.~\ref{eq_M_xeigen}, Eq.~\ref{eq_diagonalize3} and Eq.~\ref{eq_basic_force}:
\begary{rcl}
\Sigma_c&=&{1\over 4}\,Tr({\bf M})\\
\Sigma_s&=&{a\,(\alpha+\beta)+b\,(\gamma+\delta)\over 4}\\
\Delta_s&=&{a\,(\alpha+\beta)-b\,(\gamma+\delta)\over 4}\\
\endary
The last missing value to determine the coupled tunes is $\Delta_c$. We will argue
in the following, that the decoupling of ${\bf M}_s$ will casually diagonalize
${\bf M}_c$ and hence yields $\Delta_c$.

It is noteworthy that $\y_{12}$ is the only even matrix that anticommutes with all odd 
symplices (see Tab.~\ref{tab_acommtab}). This means that $\y_{12}$ plus the odd symplices 
$\y_2$, $\y_5$ and $\y_7$ build an alternative basis of $Cl(3,1)$ - in which $\y_9$ is the 
pseudoscalar (see Tab.~\ref{tab_basis_sets}).

In the following we use the diagonalization steps and reconstruct the components
of ${\bf M}_c$. The scalar part can be ignored as it can not be transformed. The
interesting part therefore is ${\bf E}\,\y_{12}\,{\bf E}^{-1}$. The matrix ${\bf E}$
is the reverse product of all transformation matrices and hence we may analyze the
structure of ${\bf M}_c$ by applying all steps of the diagonalization in reversed order.

As can be seen in Tab.~\ref{tab_commtab}, the matrix $\y_{12}$ commutes with
$\y_0$, $\y_6$ and $\y_8$, so that the reversed transformations of
Eq.~\ref{eq_diagonalize1} and Eq.~\ref{eq_diagonalize2} have no effect on $\y_{12}$.
Hence we can skip the diagonalization steps and go back to the transformations of 
Sec.~\ref{sec_decoupling_const}. The general structure of the resulting terms
can be seen from Tab.~\ref{tab_commtab}. All transformations are symplectic, i.e. are
launched by symplices. All non-vanishing commutators of RDMs in the range $\y_{10}\dots\y_{14}$ 
with symplices yield matrices in the range $\y_{10}\dots\y_{14}$. Therefore the
symplectic (back-) transformation of $\y_{12}$ exclusively yield axial-vector and
pseudoscalar components. 
In the following we assume that the one-turn-transfer matrix is known and 
``off resonance'', i.e. non of the coefficients of Eq.~\ref{eq_M_xcoeff} are zero.

Then the first step of the back-transformation is
\begary{rcl}
{\bf X}^{(1)}&=&{\bf R}_2^{-1}\,\y_{12}\,{\bf R}_2\\
             &=&\cosh{\eps_2}\,\y_{12}+\sinh{\eps_2}\,\y_{14}\\
\label{eq_x1}
\endary
The second step - the rotation about the $y$-axis using $\y_8$ again has no
effect, since both, $\y_{12}$ and $\y_{14}$ commute with $\y_8$.
Hence the next steps are the rotations about $z$- and $x$-axis.
Both rotations commute with $\y_{14}$ - so that the second term in
Eq.~\ref{eq_x1} remains unchanged. But the first term is an (axial)
vector term which rotates analogue to the vector term $\y_2$. Hence
the rotations create coefficients of $\y_{11}$ and $\y_{13}$.
The next step is the inverse lorentz boost in $y$-direction with $\y_5$. Since
$\y_5$ commutes with $\y_{11}$ and $\y_{13}$, only the $\y_{12}$-term transforms 
and launches a $\y_{10}$-component.

The remaining last two transformations are again rotations about the $x$- and $z$-axis. 
Rotations will neither change the pseudoscalar coefficient nor the time-like axial vector
component $\y_{10}$. Hence the pseudoscalar coefficient of the transfer matrix can be used 
to determine, if a phase boost is required for decoupling. According to the decoupling-recipe 
this is the case, if (and only if) the EMEQ-components of the force matrix fulfill $\vec E\vec B=0$. 

Hence one finds:
\begary{rcl}
-\sinh{\eps_2}\,\Delta_c&=&\y_{14}\cdot{\bf M}\\
{\bf\tilde X}&=&{\bf M}_c-{1\over 4}\,Tr({\bf M}_c)\\
             &=&-\Delta_c\,{\bf E}\,\y_{12}\,{\bf E}^{-1}\\
\endary
If the decoupling transformation has been computed, then
\begeq
\Delta_c=-\y_{12}\,{\bf E}^{-1}\,\left({\bf M}_c-{1\over 4}\,Tr({\bf M}_c)\,\right)\,{\bf E}\,.
\endeq
Hence we may now solve Eq.~\ref{eq_M_xcoeff} for the tunes:
\begary{rcl}
\tan{(\bar\omega\tau)}&=&{\Sigma_s\over\Sigma_c}=-{\Delta_c\over\Delta_s}\\
\tan{(\Delta\omega\tau)}&=&{\Delta_s\over\Sigma_c}={\Delta_c\over\Sigma_s}\\
\endary
If $\tau$ is the repetition time (or length of the design orbit) of the accelerator,
then the tune $Q_i$ equals $Q_i={\omega_i\tau\over 2\pi}$.

Note that the anticommutators ${\y_\mu\y_\nu+\y_\nu\y_\mu\over 2}$ for $\mu,\nu\,\in\,[10\dots 14]$ 
form a diagonal $5\times 5$-matrix and hence are the basis of a five-dimensional phase 
space (see Tab.~\ref{tab_acommtab}).

\subsection{Transformation to Equal Frequencies}
\label{sec_eqfreq}

Consider now a time dependent symplectic transformation as derived in Eq.~\ref{eq_ttrans} with 
\begeq
{\bf G}=i\,\omega\,\y_4\,,
\endeq
with the corresponding matrices 
\begary{rcl}
{\bf U}&=&e^{i\,\omega\,\y_4\,\tau}\\
       &=&\mathrm{Diag}(e^{-i\,\omega\,\tau},e^{i\,\omega\,\tau},e^{-i\,\omega\,\tau},e^{i\,\omega\,\tau})\\
{\bf U}^{-1}&=&e^{-i\,\omega\,\y_4\,\tau}\\
       &=&\mathrm{Diag}(e^{i\,\omega\,\tau},e^{-i\,\omega\,\tau},e^{i\,\omega\,\tau},e^{-i\,\omega\,\tau})\\
\endary
which does not affect the diagonalized force matrix ${\bf F}$:
\begeq
{\bf U}\,{\bf F}\,{\bf U}^{-1}={\bf F}\,,
\endeq
but results in a frequency shift such that
\begary{rcl}
{\bf F}'&=&{\bf F}+i\,\omega\,\y_4\\
        &=&\bmtx{cccc}
i\,\Omega_1&0&0&0\\
0&-i\,\Omega_1&0&0\\
0&0&i\,\Omega_2&0\\
0&0&0&-i\,\Omega_2\\
\emtx\\
\Omega_1&=&(\tilde m+B_y)\,a-\omega\\
\Omega_2&=&(\tilde m-B_y)\,b+\omega\\
\endary
One may now chose $\omega$ such that $\Omega_1=\Omega_2=\Omega$, that is: 
\begary{rcl}
(\tilde m+B_y)\,a-\omega&=&(\tilde m-B_y)\,b+\omega\\
2\,\omega&=&(\tilde m+B_y)\,a-(\tilde m-B_y)\,b\\
\omega&=&{\tilde m\,(a-b)+B_y\,(a+b)\over 2}\\
\endary
then we obtain the result that 
\begeq
{\bf F}=-i\,\Omega\,\y_3\,.
\endeq

It is also possible to chose  
\begeq
{\bf G}=i\,\omega\,\y_3\,,
\endeq
resulting in a frequency shift such that
\begary{rcl}
{\bf F}'&=&{\bf F}+i\,\omega\,\y_3\\
        &=&-i\,\Omega\,\y_4\\
\omega&=&{\tilde m\,(a+b)+B_y\,(a-b)\over 2}\\
\Omega&=&(\tilde m+B_y)\,a-\omega\,.
\endary

\subsection{Time Dependent Forces}
\label{sec_decoupling_t}

The treatment of time dependent symplectic transformations goes beyond the scope
of this article. Nevertheless we would like to make a few comments.

If the parameter $\eps$ of Eq.~\ref{eq_sine_cosine} is time dependent, it follows that
\begeq
{\bf\dot R}_b\,{\bf R}_b^{-1}=\y_b\,\dot\eps\,.
\endeq
Eq.~\ref{eq_forcetrans} can then be written as:
\begeq
{\bf \tilde F}=\y_b\,\dot\eps+{\bf R}_b\,{\bf F}\,{\bf R}_b^{-1}\,.
\label{eq_interference}
\endeq
Consider for instance the case of a force matrix ${\bf F}$ with the simple periodic form
\begeq
{\bf F}={\bf F}_0\,(1+\alpha\,\cos{(\omega\,\tau)})\,.
\endeq
Then the use of a symplectic transformation matrix 
\begeq
{\bf R}=\exp{(-\alpha/\omega\,{\bf F}_0\,\sin(\omega\,\tau))}\,,
\endeq
results in a constant force matrix:
\begary{rcl}
\dot{\tilde\psi}&=&(-\alpha\,{\bf F}_0\,\cos{(\omega\,\tau)}+{\bf F}_0+{\bf F}_0\,\alpha\,\cos{(\omega\,\tau)})\,\tilde\psi\\
                &=&{\bf F}_0\,\tilde\psi\\
\endary
The general time dependent case is more involved. If we intend to use ${\bf R}_b$ 
in order to let the corresponding odd component of the force matrix vanish, then 
Eq.~\ref{eq_interference} implies that we have to take the new (odd) term $\y_b\,\dot\eps$ 
into account.

Following the recipe from above, the first step is the alignment of the momentum
along the $y$-axis using ${\bf R}_7$ and ${\bf R}_9$, respectively.
This introduces new (time dependent) terms in the coefficients of $\y_7$ and $\y_9$ -
which is no problem at this stage. The second step is a boost along the $y$-axis
using $\y_5$ with the intention to get rid of $p_y$. Even this is not a problem,
since $\y_5$ corresponds to $E_y$, which has not yet been considered.
Hence it is always possible to transform into the ``rest frame'' at the cost of 
additional ``field'' terms in $B_x$, $B_z$ and $E_y$. 

Nevertheless the next steps include difficulties,
since the alignment of $\vec B$ along the $y$-axis induces again additional terms
in $B_x$, $B_z$. If the generators of the transformation and the terms to act
on are identical, the straightforward approach fails. The rotation about 
the $x$-axis:
\begary{rcl}
\eps&=&\arctan{\left({B_z\over B_y}\right)}\\
\dot\eps&=&{\dot B_z\,B_y-\dot B_y\,B_z\over B_y^2+B_z^2}\\
B_x&\to&B_x+{\dot B_z\,B_y-\dot B_y\,B_z\over B_y^2+B_z^2}\,.
\endary

More general one has
\begary{rcl}
\eps&=&\arctan{\left({X\over Y}\right)}\\
\Rightarrow\dot\eps&=&{\dot X\,Y-\dot Y\,X\over Y^2+X^2}\\
\eps&=&\mathrm{artanh}{\left({X\over Y}\right)}\\
\Rightarrow\dot\eps&=&{\dot X\,Y-\dot Y\,X\over Y^2-X^2}\,.
\endary

\section{Summary}

We introduced the real Dirac matrices and the concept of the symplex in
two-dimensional coupled linear optics, i.e. classical Hamiltonian mechanics. 
Since these matrices form a basis for {\it all} real-valued $4\times 4$-matrices, 
we can claim that our survey of symplectic transformations for such systems 
is complete and final. We have shown that a subset of these transformations is 
isomorphic to rotations and Lorentz boost as applied to Dirac spinors in quantum 
electrodynamics. We used this isomorphism to describe a general and 
straightforward decoupling algorithm for the case of constant forces 
and for transfer matrices in the context of periodic motion. 
The algorithm was tested numerically using random force matrices of stable
systems and it never failed in case of stable systems.

There are six possible choices for the symplectic unit matrix $\y_0$ as 
listed in Tab.~\ref{tab_basis_sets} corresponding to six antisymmetric RDMs. 
These choices are nothing but permutations of the dynamical variables in the 
state vector, and hence they do not differ in their physical content. 
For each possible $\y_0$ there are two choices for $\y_1$, $\y_2$ and $\y_3$, 
which can be transformed into each other by the phase rotation.
Hence the choice of the system of RDMs is unique as soon as
the order of the canonical coordinates and momenta is fixed.

We clarified the structure of the transfer matrix in two-dimensional coupled 
linear optics which we believe is the first universally valid generalization
of the Courant-Snyder theory. 

\section{Discussion and Outlook}

Since the Dirac matrices are usually known in the context of relativistic
quantum electrodynamics (QED), it is sort of inevitable to contemplate
the relationship between QED and classical mechanics.
It is known that the eigenvalues of stable systems lie on the unit circle in 
the complex plane~\cite{Arnold}, so that the treatment of eigenvectors - which 
are essential to the theory - requires the introduction of imaginary numbers. 
David Hestenes gave a geometrical interpretation of the unit imaginary in quantum
mechanics~\cite{Hestenes0,Hestenes1}.
In context of two-dimensional symplectic flow, the unit imaginary can be 
interpreted {\it statistically}, as the complex eigenvectors are able to replace 
an statistical ensemble of real-valued trajectories. This is to say, that 
the matrix $\sigma=\psi\bar\psi$ with a real-valued ``spinor'' $\psi$ has
a vanishing determinant, i.e. zero emittance - unless we introduce some kind
of averaging - either over a statistical ensemble or over time. The introduction
of complex-valued eigenvectors circumvents this problem as well-known in the
Courant-Snyder theory. 

There is a theorem that all symplectic systems of the same dimension are isomorphic~\cite{Arnold}. 
If the reverse is also true then the existence of the EMEQ suggests that Minkowski space-time 
is ``embedded'' in a four-dimensional phase space and that special relativity can
be ``deduced'' from classical symplectic motion. This deduction implies that the
components of the relativistic momentum are not proper canonical 
variables in the classical sense, but instead energy and momentum appear to be 
second moments of the dynamical variables of a two-dimensional symplectic system.
In fact, the invariance of mass is then formally and physically identical to the 
invariance of the phase space volume (emittance) in symplectic flow. 
This formal analogy seems to find its correspondence in the relation
\begeq
E=m\,c^2=\hbar\,\omega\,.
\endeq

There are still many open questions as for instance: why are there six field components
in electromagnetic theory instead of ten as we should expect from two-dimensional
symplectic motion? The answer could be related to the dimensionality of the
system under consideration. The phase-space dimension $n$ of a classical system 
and the number $\nu$ of ``valid'' symplices are related by Eq.~\ref{eq_dimensions}
and can be resolved for $n$:
\begeq
n=\sqrt{2\,\nu+{1\over 4}}-{1\over 2}\,.
\label{eq_dimrev}
\endeq
If we insert $\nu=6$, then we find $n=3$. Classically the phase space dimension
should be an even number, since Hamilton mechanics is based on pairs $(q_i,p_i)$.
But a generalization of Hamilton mechanics for odd phase space dimensions has already 
been presented by Nambu utilizing several distinct Hamilton functions~\cite{Nambu}.
We believe that the description of odd-dimensional phase spaces can also be done
if canonical pairs $(q_i,p_i)$ are used - if the dimensionality of a systems 
is defined by the number of available force components, i.e. the number of symplices.
Relativistic electromagnetic theory knows six symplices (and hence six symplectic
transformations), two-dimensional coupled optics has ten, in App.~\ref{sec_complex}
we argue that complexification of the presented theory leads to a $15$ valid symplices
corresponding to a $5$-dimensional phase space.

\begin{acknowledgments}
Mathematica\textsuperscript{\textregistered} $5.2$ has been used for some of the
symbolic calculations. Addition software has been written in ``C'' and been compiled 
with the GNU\textsuperscript{\copyright}-C++ compiler 3.4.6 on Scientific Linux.
\end{acknowledgments}

\begin{appendix}

\section{Complexification}
\label{sec_complex}

If we consider complex spinors then it is quickly shown that the Hamilton equations are:
\begary{rcl}
\psi_i&=&{q_i+i\,p_i\over\sqrt{2}}\\
\dot q_i&=&{i\over\sqrt{2}}\,\left({\d H\over\d\psi_i}-{\d H\over\d\psi_i^\star}\right)\\
\dot p_i&=&-{1\over\sqrt{2}}\,\left({\d H\over\d\psi_i}+{\d H\over\d\psi_i^\star}\right)\\
i\,\dot\psi_i&=&{\d H\over\d\psi_i^\star}\\
i\,\dot\psi_i^\star&=&-{\d H\over\d\psi_i}\\
\endary
Given the Hamiltonian function is expressed by
\begeq
H=\psi^\dag\,{\cal A}\,\psi=\sum\limits_{i,j}\,\psi_i^\star\,{\cal A}_{ij}\,\psi_j\,,
\endeq
with the matrix ${\cal A}$ then one obtains:
\begary{rcl}
i\,\dot\psi_k&=&{\d H\over\d\psi_k^\star}=\sum\limits_{i,j}\,\delta_{ik}\,{\cal A}_{ij}\,\psi_j\\
             &=&\sum\limits_{j}\,{\cal A}_{kj}\,\psi_j\\
i\,\dot\psi_k^\star&=&-{\d H\over\d\psi_k}=-\sum\limits_{i,j}\,\psi_i^\star\,{\cal A}_{ij}\,\delta_{jk}\\
                   &=&-\sum\limits_{i}\,\psi_i^\star\,{\cal A}_{ik}\\
\endary
With implicit summation we write
\begary{rcl}
i\,\dot\psi&=&{\cal A}\,\psi\\
i\,\dot\psi^\dag&=&-\psi^\dag\,{\cal A}\\
\endary
If we take the adjunct of the second equation
\begeq
 i\,\dot\psi=({\cal A}^T)^\star\,\psi={\cal A}^\dag\,\psi\,,
\endeq
it becomes obvious that the matrix ${\cal A}$ must be self-adjoint or {\it hermitian}. 
If we switch back to the $(q_i,p_i)$ approach with ${\cal A}={\bf A}+i\,{\bf B}$, ${\bf A}={\bf A}^T$, ${\bf B}^T=-{\bf B}$
\begary{rcl}
 H&=&\psi^\dag\,{\cal A}\,\psi\\
 H&=&{1\over 2}\,(q-i\,p)^T\,({\bf A}+i{\bf B})\,(q+i\,p)\\
 H&=&{1\over 2}\,\left( q^T\,{\bf A}\,q+p^T\,{\bf A}\,p-2\,q^T\,{\bf B}\,p\right)\,,
\label{eq_Hcomplex}
\endary
then we find additional restrictions compared to a ``classical'' system: The matrix
${\bf A}$ appears for both - coordinates and momenta. The complex notation of
a 4-dimensional spinor is therefore not suitable for the general description of 
4-dimensional classical harmonic oscillators.

The symmetric matrix ${\bf A}$ in the real-valued 2-dim. description has been replaced by an hermitian
matrix in the complex notation. It can be shown that the generalization of a symplectic transformation
matrix ${\bf U}$ then is
\begeq
\y_0={\bf U}\,\y_0\,{\bf U}^\dag\,.
\label{eq_complex_symplectic}
\endeq
There are $15$ complex Dirac matrices with zero trace that obey Eq.~\ref{eq_complex_symplectic} so that
the phase space is - according to Eq.~\ref{eq_dimrev} 5-dimensional.

\section{The $\y$-Matrices}
\label{sec_app1}

To complete the list of the real $\y$-matrices used throughout this paper:
{\small$$
\begin{array}{rclp{4mm}rcl}
\y_4&=&\bmtx{cccc}
-1& 0& 0& 0\\
 0& 1& 0& 0\\
 0& 0& 1& 0\\
 0& 0& 0&-1\\
\emtx&&
\y_5&=&\bmtx{cccc}
 0& 0& 1& 0\\
 0& 0& 0&-1\\
 1& 0& 0& 0\\
 0&-1& 0& 0\\
\emtx\\
\y_6&=&\bmtx{cccc}
 0& 1& 0& 0\\
 1& 0& 0& 0\\
 0& 0& 0& 1\\
 0& 0& 1& 0\\
\emtx&&
\y_7&=&\bmtx{cccc}
 0& 0& 0& 1\\
 0& 0&-1& 0\\
 0& 1& 0& 0\\
-1& 0& 0& 0\\
\emtx\\
\end{array}
$$}
{\small$$
\begin{array}{rclp{4mm}rcl}
\y_8&=&\bmtx{cccc}
 0& 1& 0& 0\\
-1& 0& 0& 0\\
 0& 0& 0&-1\\
 0& 0& 1& 0\\
\emtx&&
\y_9&=&\bmtx{cccc}
 0& 0&-1& 0\\
 0& 0& 0&-1\\
 1& 0& 0& 0\\
 0& 1& 0& 0\\
\emtx\\
\y_{10}&=&\bmtx{cccc}
 0& 0& 1& 0\\
 0& 0& 0&-1\\
-1& 0& 0& 0\\
 0& 1& 0& 0\\
\emtx&&
\y_{11}&=&-\bmtx{cccc}
 0& 0& 1& 0\\
 0& 0& 0& 1\\
 1& 0& 0& 0\\
 0& 1& 0& 0\\
\emtx\\
\end{array}
$$}
{\small$$
\begin{array}{rclp{4mm}rcl}
\y_{12}&=&\bmtx{cccc}
-1& 0& 0& 0\\
 0&-1& 0& 0\\
 0& 0& 1& 0\\
 0& 0& 0& 1\\
\emtx&&
\y_{13}&=&\bmtx{cccc}
 0& 0& 0&-1\\
 0& 0& 1& 0\\
 0& 1& 0& 0\\
-1& 0& 0& 0\\
\emtx\\
\y_{14}&=&\bmtx{cccc}
 0& 0& 0&-1\\
 0& 0&-1& 0\\
 0& 1& 0& 0\\
 1& 0& 0& 0\\
\emtx&&
\y_{15}&=&{\bf 1}\\
\end{array}
$$}

\begin{table}
{\tiny
\begin{tabular}{|c|c|ccc|ccc|ccc|c|ccc|c|}\hline
     &\multicolumn{4}{c|}{Basis $(\Phi,\vec A)$}&\multicolumn{3}{c|}{$\vec E$}&\multicolumn{3}{c|}{$\vec B$}&$V_0$&\multicolumn{3}{c|}{$\vec V$}& PS\\\hline
     &\multicolumn{4}{c|}{4-vector}&\multicolumn{3}{c|}{Electric Field}&\multicolumn{3}{c|}{Magnetic Field}&\multicolumn{4}{c|}{Axial 4-vector}& PS\\\hline
1)  &${0}$  &${1}$ &${2}$ &${3}$ & ${4}$ &${5}$ &${6}$ &${7}$ & ${8}$ &${9}$ &${10}$ &${11}$ & ${12}$ &${13}$ &${14}$ \\
2)  &${0}$  &${4}$ &${5}$ &${6}$ & $-{1}$ &$-{2}$ &$-{3}$ &${7}$ & ${8}$ &${9}$ &${14}$ &${11}$ & ${12}$ &${13}$ &$-{10}$ \\\hline
3)  &${14}$ &${1}$ &${2}$ &${3}$ & ${11}$ &${12}$ &${13}$ &${7}$ & ${8}$ &${9}$ &${10}$ &$-{4}$ & $-{5}$ &$-{6}$ &$-{0}$ \\
4)  &${14}$ &${11}$ &${12}$ &${13}$ & $-{1}$ &$-{2}$ &$-{3}$ &${7}$ & ${8}$ &${9}$ &$-{0}$ &$-{4}$ & $-{5}$ &$-{6}$ &$-{10}$ \\\hline
5)  &${10}$ &${11}$ &${12}$ &${13}$ & ${4}$ &${5}$ &${6}$ &${7}$ & ${8}$ &${9}$ &$-{0}$ &$-{1}$ & $-{2}$ &$-{3}$ &${14}$ \\
6)  &${10}$ &${4}$ &${5}$ &${6}$ & $-{11}$ &$-{12}$ &$-{13}$ &${7}$ & ${8}$ &${9}$ &${14}$ &$-{1}$ & $-{2}$ &$-{3}$ &${0}$ \\\hline\hline
7)  &${7}$  &${3}$ &${6}$ &${13}$ & ${2}$  &$ {5}$ &$ {12}$  &${10}$ & ${14}$ &$-{0}$ &${9}$ &$-{1}$ & $-{4}$ &$-{11}$ &${8}$ \\
8)  &${7}$  &${2}$ &${5}$ &${12}$ & $-{3}$ &$-{6}$ &$-{13}$  &${10}$ & ${14}$ &$-{0}$ &${8}$ &$-{1}$ & $-{4}$ &$-{11}$ &$-{9}$ \\\hline
9)  &${8}$  &${3}$ &${6}$ &${13}$ & $-{1}$ &$-{4}$ &$-{11}$  &${10}$ & ${14}$ &$-{0}$ &${9}$ &$-{2}$ & $-{5}$ &$-{12}$ &$-{7}$ \\
10) &${8}$  &${1}$ &${4}$ &${11}$ & ${3}$  &$ {6}$ &$ {13}$  &${10}$ & ${14}$ &$-{0}$ &${7}$ &$-{2}$ & $-{5}$ &$-{12}$ &${9}$ \\\hline
11) &${9}$  &${1}$ &${4}$ &${11}$ & $-{2}$ &$-{5}$ &$-{12}$  &${10}$ & ${14}$ &$-{0}$ &${7}$ &$-{3}$ & $-{6}$ &$-{13}$ &$-{8}$ \\
12) &${9}$  &${2}$ &${5}$ &${12}$ & ${1}$  &$ {4}$ &$ {11}$  &${10}$ & ${14}$ &$-{0}$ &${8}$ &$-{3}$ & $-{6}$ &$-{13}$ &${7}$ \\\hline
\end{tabular}}
\caption{List of the indices of equivalent canonical basis sets of $\y$-matrices and the 
corresponding re-interpretations of the matrices relative to the choice
of the basis. Negative indices indicate the negative matrix with the index taken as absolute value. 
$(V_0,\vec V)$ are the elements corresponding to the so--called ``axial vector'', ``PS'' is the pseudo-scalar.
\label{tab_basis_sets}}
\end{table}

\section{Arbitrary Hamiltonians}
\label{sec_app2}

The concept presented above is far more general, as any {\it arbitrary} Hamiltonian $H(\Psi)$ 
may be written as a Taylor serie up to second order:
\begeq
H(\Psi)=H_0+\sum\limits_k\,\eps_k\,\Psi_k+{1\over 2}\,\sum\limits_{i,k}\,\Psi_i\,A_{ik}\,\Psi_k+\dots\,.
\label{eq_HTaylor}
\endeq
From Eq.~\ref{eq_HTaylor} and Eq.~\ref{eq_eqom_general} one finds the following EQOM:
\begeq
\dot\Psi=\y_0\,\eps+\y_0\,{\bf A}\,\Psi\,.
\label{eq_symmAi}
\endeq 
Leach formulated a time canonical transformation to transform a time-dependent
Hamiltonian of the form given by EQ.~\ref{eq_HTaylor} to a different (simpler)
form, which we will summarize in the following. Given a symplectic time-dependent 
transformation matrix ${\bf R}$, one writes for the spinor $\psi$ in the transformed system:
\begeq
\psi={\bf R}\,\Psi+\psi_0\,.
\label{eq_psi_trans}
\endeq
The new Hamiltonian $\tilde H$ should have the form
\begeq
\tilde H=\tilde H_0+\tilde\eps\,\psi+{1\over 2}\,\psi^T\,\tilde{\bf A}\,\psi\,.
\label{eq_HTaylor2}
\endeq
and the new EQOM are
\begeq
\dot{\psi}=\y_0\,\tilde\eps+\y_0\tilde {\bf A}\,\psi\,,
\label{eq_eqom_trans}
\endeq
or - using EQ.~\ref{eq_psi_trans} and EQ.~\ref{eq_symmAi}:
\begeq
\dot\psi={\bf \dot R}\,\Psi+{\bf R}\,\dot\Psi+\dot\psi_0={\bf \dot R}\,\Psi+{\bf R}\,(\y_0\,\eps+\y_0\,{\bf A}\,\Psi)+\dot\psi_0\,.
\label{eq_psi_trans2}
\endeq
Combining EQ.~\ref{eq_eqom_trans} and EQ.~\ref{eq_psi_trans2} yields:
\begary{rcl}
\dot\psi_0&=&\y_0\,\tilde {\bf A}\,\psi_0-{\bf R}\,\y_0\,\eps+\y_0\,\tilde\eps\\
{\bf \dot R}&=&\y_0\,\tilde{\bf A}\,{\bf R}-{\bf R}\,\y_0\,{\bf A}\,,
\label{eq_Sdot}
\endary
where the ``coordinate-free and coordinate-dependent parts have been separated''~\cite{Leach}. 

We are especially interested in a transformation which keeps the forces 
$\y_0\,{\bf A}=\y_0\,\tilde{\bf A}$ but sets $\tilde\eps=0$ in the new system.
One obtains with ${\bf F}=\y_0\,{\bf A}$:
\begary{rcl}
\dot{\psi}&=&{\bf F}\,\psi\\
\dot\psi_0&=&{\bf F}\,\psi_0-{\bf R}\,\y_0\,\eps\\
{\bf \dot R}&=&{\bf F}\,{\bf R}-{\bf R}\,{\bf F}\,,
\label{eq_SdotX}
\endary
If one considers the (rather trivial) situation that $\eps=0$, then the conclusion
is that any spinor $\psi$ can formally be replaced by $\psi_0+{\bf R}\,\psi$ -
or more generally: Given a set of vectors $\psi_k$ where each vector $\psi_k$ 
fulfills the EQOM 
\begeq
\dot\psi_k={\bf F}\,\psi_k
\endeq
then one finds that a superposition of the form
\begeq
\psi=\sum\limits_k\,{\bf R}_k\,\psi_k
\endeq
where the symplectic matrices ${\bf R}_k$ obey the EQOM
\begeq
\dot{\bf R}_k={\bf F}\,{\bf R}_k-{\bf R}_k\,{\bf F}\,,
\endeq
also fulfills the EQOM:
\begary{rcl}
\dot\psi&=&\sum\limits_k\,{\bf R}_k\,\dot\psi_k+\sum\limits_k\,\dot{\bf R}_k\,\psi_k\\
        &=&\sum\limits_k\,{\bf R}_k\,{\bf F}\,\psi_k+\sum\limits_k\,\left({\bf F}\,{\bf R}_k-{\bf R}_k\,{\bf F}\,\right)\,\psi_k\\
        &=&{\bf F}\,\sum\limits_k\,{\bf R}_k\,\psi_k\\
        &=&{\bf F}\,\psi\\
\endary

\section{Multiplication Tables of the RDMs}
\label{sec_app3}

Tab.~\ref{tab_commtab} lists the commutators and Tab.~\ref{tab_acommtab} the anticommutators of all RDMs.
The sum of both is the multiplication table of the group.
\begin{table}
{\tiny\begin{tabular}{|c||c|c|c|c|c|c|c|c|c|c|c|c|c|c|c|c|c|}\hline
     &    0&    1&    2&    3&    4&    5&    6&    7&    8&    9&   10&   11&   12&   13&   14&   15 \\\hline\hline
    0&     &    4&    5&    6&   -1&   -2&   -3&     &     &     &   14&     &     &     &  -10&     \\\hline
    1&   -4&     &    9&   -8& -0 &     &     &     &   -3&    2&     &  -14&     &     &  -11&     \\\hline
    2&   -5&   -9&     &    7&     & -0 &     &    3&     &   -1&     &     &  -14&     &  -12&     \\\hline
    3&   -6&    8&   -7&     &     &     & -0 &   -2&    1&     &     &     &     &  -14&  -13&     \\\hline
    4&    1&    0&     &     &     &    9&   -8&     &   -6&    5&   11&   10&     &     &     &     \\\hline
    5&    2&     &    0&     &   -9&     &    7&    6&     &   -4&   12&     &   10&     &     &     \\\hline
    6&    3&     &     &    0&    8&   -7&     &   -5&    4&     &   13&     &     &   10&     &     \\\hline
    7&     &     &   -3&    2&     &   -6&    5&     &   -9&    8&     &     &  -13&   12&     &     \\\hline
    8&     &    3&     &   -1&    6&     &   -4&    9&     &   -7&     &   13&     &  -11&     &     \\\hline
    9&     &   -2&    1&     &   -5&    4&     &   -8&    7&     &     &  -12&   11&     &     &     \\\hline
   10&  -14&     &     &     &  -11&  -12&  -13&     &     &     &     &    4&    5&    6&    0&     \\\hline
   11&     &   14&     &     &  -10&     &     &     &  -13&   12&   -4&     &    9&   -8&    1&     \\\hline
   12&     &     &   14&     &     &  -10&     &   13&     &  -11&   -5&   -9&     &    7&    2&     \\\hline
   13&     &     &     &   14&     &     &  -10&  -12&   11&     &   -6&    8&   -7&     &    3&     \\\hline
   14&   10&   11&   12&   13&     &     &     &     &     &     & -0 &   -1&   -2&   -3&     &     \\\hline
   15&     &     &     &     &     &     &     &     &     &     &     &     &     &     &     &     \\\hline
\end{tabular}}
\caption{Commutator table of the group of $\y$-matrices: $\y_c={\y_a\,\y_b-\y_b\,\y_a\over 2}$ where $a$ is the index
of the row, $b$ the index of the column and $c$ the table entry. The only matrix commuting with all other matrices,
is the unit matrix $\y_{15}={\bf 1}$
\label{tab_commtab}}
\end{table}

\begin{table}
{\tiny\begin{tabular}{|c||c|c|c|c|c|c|c|c|c|c|c|c|c|c|c|c|c|}\hline
     &    0&    1&    2&    3&    4&    5&    6&    7&    8&    9&   10&   11&   12&   13&   14&   15 \\\hline\hline
    0&  -15&     &     &     &     &     &     &   11&   12&   13&     &   -7&   -8&   -9&     &    0 \\\hline
    1&     &   15&     &     &     &  -13&   12&   10&     &     &    7&     &    6&   -5&     &    1 \\\hline
    2&     &     &   15&     &   13&     &  -11&     &   10&     &    8&   -6&     &    4&     &    2 \\\hline
    3&     &     &     &   15&  -12&   11&     &     &     &   10&    9&    5&   -4&     &     &    3 \\\hline
    4&     &     &   13&  -12&   15&     &     &   14&     &     &     &     &   -3&    2&    7&    4 \\\hline
    5&     &  -13&     &   11&     &   15&     &     &   14&     &     &    3&     &   -1&    8&    5 \\\hline
    6&     &   12&  -11&     &     &     &   15&     &     &   14&     &   -2&    1&     &    9&    6 \\\hline
    7&   11&   10&     &     &   14&     &     &  -15&     &     &   -1& -0 &     &     &   -4&    7 \\\hline
    8&   12&     &   10&     &     &   14&     &     &  -15&     &   -2&     & -0 &     &   -5&    8 \\\hline
    9&   13&     &     &   10&     &     &   14&     &     &  -15&   -3&     &     & -0 &   -6&    9 \\\hline
   10&     &    7&    8&    9&     &     &     &   -1&   -2&   -3&  -15&     &     &     &     &   10 \\\hline
   11&   -7&     &   -6&    5&     &    3&   -2& -0 &     &     &     &   15&     &     &     &   11 \\\hline
   12&   -8&    6&     &   -4&   -3&     &    1&     & -0 &     &     &     &   15&     &     &   12 \\\hline
   13&   -9&   -5&    4&     &    2&   -1&     &     &     & -0 &     &     &     &   15&     &   13 \\\hline
   14&     &     &     &     &    7&    8&    9&   -4&   -5&   -6&     &     &     &     &  -15&   14 \\\hline
   15&    0&    1&    2&    3&    4&    5&    6&    7&    8&    9&   10&   11&   12&   13&   14&   15 \\\hline
\end{tabular}}
\caption{Anticommutator-table of the group of $\y$-matrices: $\y_c={\y_a\,\y_b+\y_b\,\y_a\over 2}$ where $a$ is the index
of the row, $b$ the index of the column and $c$ the table entry. 
\label{tab_acommtab}}
\end{table}

\section{Second Moments and Expectation Values}
\label{sec_app4}

In order to obtain the direct relation between the second moments $\sigma_{ij}$ 
according to Eq.~\ref{eq_sigma_def} and the expectation values $\tilde f_k$, we
established two new vectors. The first is a vector of second moments:
\begeq
\vec\sigma=(\sigma_{11},\sigma_{22},\sigma_{33},\sigma_{44},\sigma_{12},\sigma_{34},\sigma_{13},\sigma_{24},\sigma_{23},\sigma_{14})^T\,,
\endeq
and the second is the vector $\vec f$ of the form
\begeq
\vec x=(\tilde f_0,\tilde f_1,\tilde f_6,\tilde f_8,\tilde f_3,\tilde f_4,\tilde f_2,\tilde f_7,\tilde f_5,\tilde f_9)^T\,.
\endeq
Ordered in this form, the equation
\begeq
\vec\sigma={\bf T}\,\vec x
\endeq
holds for a block-diagonal matrix ${\bf T}$:
{\small\begary{rcl}
{\bf T}&=&{1\over 2}\bmtx{cccccccccc}
  1 &-1 & 1 & 1 & &&&&&\\
  1 & 1 &-1 & 1 & &&&&&\\
  1 & 1 & 1 &-1 & &&&&&\\
  1 &-1 &-1 &-1 & &&&&&\\
  & &  &  & 1& 1 &&&&\\
  & &  &  & 1&-1 &&&&\\
&&  & &  &  & 1& 1 &&\\
&&  & &  &  &-1& 1 &&\\
&&&&  & &  &  & -1& -1 \\
&&&&  & &  &  & -1&  1 \\
\emtx\\
\label{eq_f2sig}
\endary}
Note that the ordering is chosen such that the first six (last four) elements of $\vec x$ 
correspond to expectation values of even (odd) RDMs. The inverse ${\bf T}^{-1}$ has the
same block-structure.

\end{appendix}

\end{document}